\documentclass[fleqn,usenatbib]{mnras}

\usepackage{graphicx}	
\usepackage{amsmath}	
\usepackage{amssymb}	
\usepackage{bm}		

\usepackage{newtxtext,newtxmath}

\usepackage[T1]{fontenc}
\usepackage{ae,aecompl}

\usepackage[usenames,dvipsnames]{xcolor}

\newcommand{\lya}{Ly$\alpha$}

\newcommand{\hi}{\ion{H}{i}}
\newcommand{\kms}{km~s$^{-1}$}

\newcommand{\hinvMpc}{$h^{-1}\,$Mpc}
\newcommand{\zforest}{$z_{\rm forest}$}

\newcommand{\mfp}{$\lambda_{\rm mfp}$}
\newcommand{\mfpz}{$\lambda_{\rm mfp}(z)$}

\newcommand{\cii}{C~{\sc ii}}
\newcommand{\kappabg}{$\kappa^{\rm bg}_{912}$}
\newcommand{\Req}{$R_{\rm eq}$}

\title[Mean free path at $z=5$--6]{The mean free path of ionizing photons at 5 <  {\emph z} < 6: evidence for rapid evolution near reionization}

\author[G. D. Becker et al.]{George D. Becker,$^1$\thanks{E-mail: george.becker@ucr.edu}
Anson D'Aloisio,$^1$ 
Holly M. Christenson,$^1$ 
Yongda Zhu,$^1$ \newauthor
G\'{a}bor Worseck$^2$ and 
James S. Bolton$^3$
\\
$^1$Department of Physics \& Astronomy, University of California, Riverside, CA, 92521, USA \\
$^2$Institut f\"{u}r Physik und Astronomie, Universit\"{a}t Potsdam, Karl-Liebknecht-Str. 24/25, D-14476 Potsdam, Germany \\
$^3$School of Physics and Astronomy, University of Nottingham, University Park, Nottingham NG7 2RD, UK
}

\date{Accepted 2021 September 14. Received 2021 September 1; in original form 2021 April 2.}

\pubyear{2021}

\begin{document}
\label{firstpage}
\pagerange{\pageref{firstpage}--\pageref{lastpage}}
\maketitle

\begin{abstract}
The mean free path of ionizing photons, \mfp, is a key factor in the photoionization of the intergalactic medium (IGM).  At $z \gtrsim 5$, however, \mfp\ may be short enough that measurements towards QSOs are biased by the QSO proximity effect.  We present new direct measurements of \mfp\ that address this bias and extend up to $z \sim 6$ for the first time.  Our measurements at $z \sim 5$ are based on data from the Giant Gemini GMOS survey and new Keck LRIS observations of low-luminosity QSOs.  At $z \sim 6$ we use QSO spectra from Keck ESI and VLT X-Shooter.  We measure $\lambda_{\rm mfp} = 9.09^{+1.62}_{-1.28}$ proper Mpc and $0.75^{+0.65}_{-0.45}$ proper Mpc (68\% confidence) at $z = 5.1$ and 6.0, respectively.  The results at $z = 5.1$ are consistent with existing measurements, suggesting that bias from the proximity effect is minor at this redshift.  At $z = 6.0$, however, we find that neglecting the proximity effect biases the result high by a factor of two or more.  Our measurement at $z = 6.0$ falls well below extrapolations from lower redshifts, indicating rapid evolution in \mfp\ over $5 < z < 6$.  This evolution disfavors models in which reionization ended early enough that the IGM had time to fully relax hydrodynamically by $z = 6$, but is qualitatively consistent with models wherein reionization completed at $z = 6$ or even significantly later.  Our mean free path results are most consistent with late reionization models wherein the IGM is still 20\% neutral at $z=6$, although our measurement at $z = 6.0$ is even lower than these models prefer.

\end{abstract}

\begin{keywords}
intergalactic medium - quasars: absorption lines - cosmology: observations - dark ages, reionization, first stars - large-scale structure of the Universe
\end{keywords}

\section{Introduction} \label{sec:intro}

The metagalactic UV background is a fundamental link between the intergalactic medium (IGM) and the sources of ionizing radiation (stars and active galactic  nuclei).  Much of our knowledge of the IGM comes from observations of the \lya\ forest, whose opacity depends directly on the hydrogen ionization rate, $\Gamma$.  For a given ionizing emissivity, $\epsilon$, the ionization rate scales roughly as $\Gamma \propto \epsilon \lambda_{\rm mfp}$ \citep[e.g.,][]{haardt2012}, where \mfp\ is the mean free path of ionizing photons.  Accurate measurements of \mfp\ are therefore essential for translating the measured properties of the IGM into constraints on the ionizing sources. 

The redshift evolution of \mfp\ may also reflect the timing of reionization \citep[e.g.,][]{rahmati2018}.  A number of observations now suggest that reionization had a midpoint around $z \sim $7--8 and ended near $z \sim 6$, or even later.  These include (i) the electron optical depth to CMB photons \citep{planckcollaboration2020}, (ii) the decline in \lya\ emission from galaxies at $z > 6$  \citep[e.g.,][and references therein]{jung2020,morales2021}, (iii)  large-scale opacity fluctuations in the \lya\ forest at $z < 6$ \citep{fan2006a,becker2015,bosman2018,eilers2018,yang2020}, (iv) the association of large \lya\ troughs at $z \sim 5.7$ with galaxy underdensities \citep{becker2018,kashino2020}, (v) \lya\ damping wings seen in the spectra of $z \sim 7$ QSOs \citep{mortlock2011,greig2017,greig2019,davies2018a,banados2018,wang2020}, (vi) the thermal history of the IGM at $z > 5$ \citep{boera2019,walther2019,gaikwad2020}, and (vii) the evolution in the number density of neutral metal absorbers near $z \sim 6$ \citep{becker2011,becker2019,cooper2019,doughty2019}.  If reionization did end near or below $z = 6$, then the mean free path at $z < 6$ should increase rapidly with time as large \ion{H}{ii} bubbles merge and the last remaining neutral islands are ionized \citep[e.g.,][]{wyithe2008}. Indeed, recent models of late reionization exhibit a rapid evolution in \mfp\ over $5 < z < 6$ \citep[e.g.,][]{kulkarni2019,keating2020,keating2020a,cain2021}.  Additionally, absorbers in recently reionized gas are photoevaporated or pressure smoothed over a time scale $\Delta t \sim 100$ Myr, contributing further to the rapid evolution in LyC opacity.  In constrast, a significantly earlier reionization would give the IGM more time to relax hydrodynamically, producing a more gradual evolution in \mfp\ at $z < 6$ \citep{park2016,daloisio2020,cain2021}.

Multiple techniques have been used to measure the mean free path.  One approach is to calculate the ionizing opacity from the incidence rate of individual \hi\ absorbers \citep[e.g.,][]{miralda-escude1990,meiksin1993,haardt1996,faucher-giguere2008b,songaila2010,rudie2013,prochaska2014}.  Alternatively, one may directly estimate the opacity from the shape of the transmitted flux profile blueward of the Lyman limit in the mean spectra of QSOs \citep{prochaska2009}.  The latter approach has arguably produced the most precise estimates of the mean free path at high redshifts, with results now spanning $2 \lesssim z \lesssim 5$ \citep{prochaska2009,fumagalli2013,omeara2013,worseck2014,lusso2018}.  One can also infer the mean free path from the average of free paths along individual QSO lines of sight \citep{romano2019}.

One challenge at $z > 5$ is that the mean free path may be comparable to or shorter than the typical size of a QSO proximity zone.  In that case, the ionizing flux from a QSO will tend to decrease the opacity in its vicinity, leading to mean free path measurements based on QSO spectra that are biased high \citep[e.g.,][]{worseck2014,daloisio2018}.  One possible solution is to use fainter QSOs, for which the impact of the proximity zone will be decreased \citep[see discussion in][]{worseck2014}.  This is observationally challenging, however, particularly given the low levels of transmission expected at $z \gtrsim 5.5$.  At $z \sim 6$ it is currently impractical to obtain enough high-quality spectra of QSOs that are sufficiently faint to meaningfully avoid the proximity effect. 

In this work we perform new measurements of the mean free path at $z > 5$, including the first direct measurement at $z \sim 6$, that address the proximity effect in two ways.  First, we modify the direct measurement approach of \citet{prochaska2009} to include a scaling of the opacity with the local ionization rate.  This allows us to account for the decrease in opacity in the vicinity of a QSO.  Second, we measure \mfp\ at $z \simeq 5$ from two groups of QSOs spanning a factor of five in mean luminosity.  This provides additional leverage in separating the background opacity from the impact of the QSOs.

The rest of the paper is organized as follows.  We describe the individual QSO spectra and the composites in Section~\ref{sec:data}.  In Section~\ref{sec:measurements} we outline our model formalism, perform tests with mock spectra, and derive measurements of \mfp.  We then discuss the implications of our results for reionization in Section~\ref{sec:discussion} before summarizing the results in Section~\ref{sec:summary}.  Our observational results assume a $\Lambda$CDM cosmology with $(\Omega_{\rm m}, \Omega_{\Lambda}, H_0) = (0.3,0.7,70~{\rm km\;s^{-1}\;Mpc^{-1}})$.  Distances are quoted in proper Mpc (pMpc) except where noted.

\section{Data}\label{sec:data}

\subsection{Samples}\label{sec:samples}

\begin{figure}
   \begin{center}
   \includegraphics[width=0.45\textwidth]{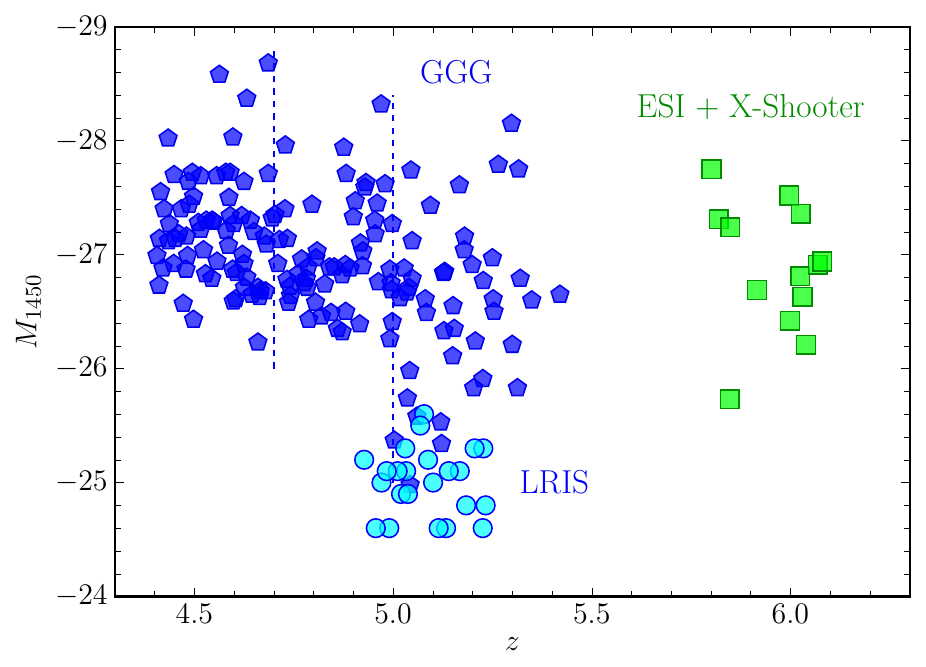}
   \vspace{-0.05in}
   \caption{Absolute magnitude at rest-frame 1450~\AA\ versus redshift for the QSOs in this work.  The full GGG sample from \citet{worseck2014} is shown for reference (dark blue pentagons) with vertical dashed lines marking their redshift bins.  In this work we analyze the GGG QSOs at $z > 5$, along with the samples observed with LRIS (light blue circles) and ESI + X-Shooter (green squares).  Details of the individual QSOs are given in Table~\ref{tab:qsos}.}
   \label{fig:M1450}
   \end{center}
\end{figure}

This work uses spectra from three QSO samples.  First, we use a subset of the spectra from the Giant Gemini GMOS (GGG) survey presented by \citet{worseck2014}.  Specifically, we use 40 QSOs spanning $5.00 < z < 5.42$, which have a mean redshift of $\langle z \rangle = 5.16$ and an absolute magnitude corresponding to the mean luminosity at rest-frame 1450~\AA\ of $M_{1450} = -26.8$.  Second, we include a sample of lower-luminosity QSOs at $z \sim 5$ observed with the Keck LRIS spectrograph.  The LRIS sample includes 23 QSOs spanning $4.93 < z < 5.24$ with $\langle z \rangle = 5.09$ and an absolute magnitude corresponding to the mean luminosity of $M_{1450} = -25.1$.  This is a factor of five fainter than the GGG sample.  Finally, we use a sample of 13 QSOs at $z \sim 6$ observed with the Keck ESI and VLT X-Shooter spectrographs.  This sample spans $5.82 < z < 6.08$ with $\langle z \rangle = 5.97$ and an absolute magnitude corresponding to the mean luminosity of $M_{1450} = -27.0$.  The samples are summarized in Table~\ref{tab:samples}, and the QSOs included in each sample are listed in Table~\ref{tab:qsos}.  We plot the rest-frame 1450~\AA\ absolute magnitudes as a function of redshift for all of our QSOs in Fig.~\ref{fig:M1450}.  

\begin{table}
   \caption{QSO samples analyzed in this work}
   \label{tab:samples}
   \begin{center}
   \begin{tabular*}{8.4cm}{@{\extracolsep{\fill}}lccc}
   \hline
   Sample  &  $n_{\rm qso}$  &  $\langle z_{\rm qso} \rangle$  &  $M_{1450}$ \\
   \hline
   GGG              &  40  &  5.16  &  $-26.6$ \\
   LRIS             &  23  &  5.09  &  $-25.1$  \\
   ESI + X-Shooter  &  13  &  5.97  &  $-27.0$  \\  
   \hline
   \end{tabular*}
   \begin{flushleft}
   For we each sample we list the number of QSOs, the mean redshift, and the absolute magnitude at rest-frame 1450~\AA\ corresponding to the mean luminosity.
   \end{flushleft}
   \end{center}
\end{table}

\begin{table}
   \caption{QSOs analyzed in this work}
   \label{tab:qsos}
   \begin{center}
   \begin{tabular*}{8.4cm}{@{\extracolsep{\fill}}lccc}
   \hline
   QSO  &  Instrument  &  $z_{\rm qso}^{a}$  &  $M_{1450}^{b}$ \\
   \hline
   SDSS J0231$-$0728  &  GMOS  &  5.420  &  $-$26.6  \\
   SDSS J0338$+$0021  &  GMOS  &  5.040  &  $-$26.7  \\
   SDSS J0824$+$1302  &  GMOS  &  5.207  &  $-$26.2  \\
   SDSS J0846$+$0800  &  GMOS  &  5.028  &  $-$26.9  \\
   SDSS J0854$+$2056  &  GMOS  &  5.179  &  $-$27.0  \\
   SDSS J0902$+$0851  &  GMOS  &  5.226  &  $-$25.9  \\
   SDSS J0913$+$5919  &  GMOS  &  5.122  &  $-$25.3  \\
   SDSS J0915$+$4924  &  GMOS  &  5.199  &  $-$26.9  \\
   SDSS J0922$+$2653  &  GMOS  &  5.042  &  $-$26.0  \\
   SDSS J0957$+$0610  &  GMOS  &  5.167  &  $-$27.6  \\
   SDSS J1026$+$2542  &  GMOS  &  5.254  &  $-$26.5  \\
   SDSS J1050$+$5804  &  GMOS  &  5.151  &  $-$26.5  \\
   SDSS J1053$+$5804  &  GMOS  &  5.250  &  $-$27.0  \\
   SDSS J1054$+$1633  &  GMOS  &  5.154  &  $-$26.4  \\
   SDSS J1101$+$0531  &  GMOS  &  5.045  &  $-$27.7  \\
   SDSS J1132$+$1209  &  GMOS  &  5.180  &  $-$27.2  \\
   SDSS J1148$+$3020  &  GMOS  &  5.128  &  $-$26.3  \\
   SDSS J1154$+$1341  &  GMOS  &  5.060  &  $-$25.6  \\
   SDSS J1202$+$3235  &  GMOS  &  5.298  &  $-$28.1  \\
   SDSS J1204$-$0021  &  GMOS  &  5.094  &  $-$27.4  \\
   SDSS J1209$+$1831  &  GMOS  &  5.127  &  $-$26.8  \\
   SDSS J1221$+$4445  &  GMOS  &  5.203  &  $-$25.8  \\
   SDSS J1222$+$1958  &  GMOS  &  5.120  &  $-$25.5  \\
   SDSS J1233$+$0622  &  GMOS  &  5.300  &  $-$26.2  \\
   SDSS J1242$+$5213  &  GMOS  &  5.036  &  $-$25.7  \\
   SDSS J1334$+$1220  &  GMOS  &  5.130  &  $-$26.8  \\
   SDSS J1337$+$4155  &  GMOS  &  5.018  &  $-$26.6  \\
   SDSS J1340$+$2813  &  GMOS  &  5.349  &  $-$26.6  \\
   SDSS J1340$+$3926  &  GMOS  &  5.048  &  $-$26.8  \\
   SDSS J1341$+$3510  &  GMOS  &  5.252  &  $-$26.6  \\
   SDSS J1341$+$4611  &  GMOS  &  5.003  &  $-$25.4  \\
   SDSS J1423$+$1303  &  GMOS  &  5.048  &  $-$27.1  \\
   SDSS J1436$+$2132  &  GMOS  &  5.227  &  $-$26.8  \\
   SDSS J1437$+$2323  &  GMOS  &  5.320  &  $-$26.8  \\
   SDSS J1534$+$1327  &  GMOS  &  5.043  &  $-$25.0  \\
   SDSS J1614$+$2059  &  GMOS  &  5.081  &  $-$26.6  \\
   SDSS J1614$+$4640  &  GMOS  &  5.313  &  $-$25.8  \\
   SDSS J1626$+$2751  &  GMOS  &  5.265  &  $-$27.8  \\
   SDSS J1659$+$2709  &  GMOS  &  5.316  &  $-$27.7  \\
   SDSS J2228$-$0757  &  GMOS  &  5.150  &  $-$26.1  \\
   J0015$-$0049       &  LRIS  &  4.931  &  $-$25.2  \\
   J0023$-$0018       &  LRIS  &  5.037  &  $-$25.1  \\
   J0108$-$0100       &  LRIS  &  5.118  &  $-$24.6  \\
   J0115$+$0015       &  LRIS  &  5.144  &  $-$25.1  \\
   J0129$-$0028       &  LRIS  &  5.015  &  $-$25.1  \\
   J0208$-$0112       &  LRIS  &  5.231  &  $-$25.3  \\
   J0221$-$0342       &  LRIS  &  5.024  &  $-$24.9  \\
   J0236$-$0108       &  LRIS  &  4.974  &  $-$25.0  \\
   J0256$+$0002       &  LRIS  &  4.960  &  $-$24.6  \\
   J0321$+$0029       &  LRIS  &  5.041  &  $-$24.9  \\
   J0338$+$0018       &  LRIS  &  4.988  &  $-$25.1  \\
   J0349$+$0034       &  LRIS  &  5.209  &  $-$25.3  \\
   J1408$+$5300       &  LRIS  &  5.072  &  $-$25.5  \\
   J1414$+$5732       &  LRIS  &  5.188  &  $-$24.8  \\
   J2111$+$0053       &  LRIS  &  5.034  &  $-$25.3  \\
   J2202$+$0131       &  LRIS  &  5.229  &  $-$24.6  \\
   J2211$+$0011       &  LRIS  &  5.237  &  $-$24.8  \\
   J2226$-$0109       &  LRIS  &  4.994  &  $-$24.6  \\
   J2233$-$0107       &  LRIS  &  5.104  &  $-$25.0  \\
   J2238$-$0027       &  LRIS  &  5.172  &  $-$25.1  \\
   J2239$+$0030       &  LRIS  &  5.092  &  $-$25.2  \\
   J2312$+$0100       &  LRIS  &  5.082  &  $-$25.6  \\
   J2334$-$0010       &  LRIS  &  5.137  &  $-$24.6  \\
   \hline
   \end{tabular*}
   \end{center}
\end{table}

\addtocounter{table}{-1}

\begin{table}
   \caption{-- {\it continued}}
   \label{tab:qsos_cont}
   \begin{center}
   \begin{tabular*}{8.4cm}{@{\extracolsep{\fill}}lccc}
   \hline
   QSO  &  Instrument  &  $z_{\rm qso}^{a}$  &  $M_{1450}$ \\
   \hline
   SDSS J0002$+$2550   &  ESI        &  5.824  &  $-$27.3  \\
   SDSS J0005$-$0006   &  ESI        &  5.851  &  $-$25.7  \\
   SDSS J0818$+$1722   &  X-Shooter  &  6.001  &  $-$27.5  \\
   SDSS J0836$+$0054   &  X-Shooter  &  5.805  &  $-$27.8  \\
   SDSS J0840$+$5624   &  ESI        &  5.853  &  $-$27.2  \\
   SDSS J0842$+$1218   &  X-Shooter  &  6.0754$^d$  &  $-$26.9  \\
   SDSS J1137$+$3549   &  ESI        &  6.030  &  $-$27.4  \\
   ULAS J1207$+$0630   &  X-Shooter  &  6.0366$^c$  &  $-$26.6  \\
   SDSS J1306$+$0356   &  X-Shooter  &  6.0330$^d$  &  $-$26.8  \\
   SDSS J1411$+$1217   &  ESI        &  5.920  &  $-$26.7  \\
   SDSS J1602$+$4228   &  ESI        &  6.084  &  $-$26.9  \\
   SDSS J2054$-$0005   &  ESI        &  6.0389$^d$  &  $-$26.2  \\
   PSO J340$-$18       &  X-Shooter  &  6.0007$^e$  &  $-$26.4  \\
   \hline
   \end{tabular*}
   \begin{flushleft}
   $^a$Redshifts for GMOS QSOs are adopted from \cite{worseck2014}.  Other redshifts quoted to three decimal places are based on the apparent start of the \lya\ forest.  See text for details. \\
   $^b$$M_{1450}$ values for GMOS QSOs were calculated from the flux-calibrated spectra published by \citet{worseck2014}.  For LRIS QSOs they are adopted from \citet{mcgreer2013,mcgreer2018}.  For ESI and X-Shooter QSOs the $M_{1450}$ values are from \citet{banados2016} and references therein. \\
   $^c$[\cii]~158~$\mu$m redshift from \citet{decarli2018} \\
   $^d$[\cii]~158~$\mu$m redshift from \citet{venemans2020} \\
   $^e$\lya\ halo redshift from \citet{farina2019} \\
   \end{flushleft}
   \end{center}
\end{table}

\subsection{GGG spectra}

Our subset of the GGG data includes all objects at $z > 5$ observed in that survey apart from one flagged as a broad absorption line (BAL) QSO and one whose flux was affected by very poor sky subtraction.  The 40 QSOs selected are listed in Table~\ref{tab:qsos}.  Details of the observation and data reduction are given in \citet{worseck2014}.  Here we note that the Lyman continuum portion of the spectra were observed with the GMOS B600 grating through a 1'' slit, which gives a FWHM resolution of roughly 320 \kms.  We also note that the spectra contain noticeable variations in the sky-level zero point, as discussed by \citet{worseck2014}.  We account for these variations when fitting models to the mean flux profile (see Section~\ref{sec:measurements}).

\subsection{LRIS observations}\label{sec:lris}

We observed 27 faint ($M_{\rm 1450} \sim -25$) $z \sim 5$ QSOs in March and September 2019 using the Keck Low Resolution Imaging Spectrometer \citep[LRIS;][]{oke1995}.  The targets were drawn from the surveys for faint QSOs conducted by \citet{mcgreer2013,mcgreer2018} in the SDSS Stripe 82 and the CFHT Legacy Survey fields.  We used a $1.0\arcsec$ slit with the D680 dichroic.  On the blue side we used the 300/5000 grism, which provided the maximum sensitivity near the Lyman limit for the QSOs in our sample (observed wavelengths near 5400--5700~\AA).  The resolution from this grism is relatively low (${\rm FWHM} \simeq 490$~\kms, measured from skylines) but sufficient for the mean free path measurement described in Section~\ref{sec:measurements}.  On the red side we used the 831/8200 grating (${\rm FWHM} \simeq 110$~\kms) centered at 7989~\AA, which allowed us to identify individual absorption lines near the start of the \lya\ forest.

The spectra were reduced using a custom reduction package similar to the one described in \citet{becker2012} and \citet{lopez2016}.  Individual frames were sky-subtracted using an optimal algorithm based on \citet{kelson2003}.  Preliminary one-dimensional spectra were then optimally extracted following \citet{horne1986} .  For each exposure, a telluric absorption model was fit to the red side and then propagated back to the two-dimensional sky-subtracted frames for both the blue and the red side.    A final one-dimensional spectrum for each side was then extracted simultaneously from all exposures of a given object.  One complication of our chosen setup is that the D680 dichroic combined with the 300/5000 grism allows contamination from second-order light.  This is nominally not a problem for our QSOs, which have essentially no flux blueward of $\sim$5000~\AA; however, it does impact the spectra of blue standard stars, which in turn can impact the flux calibration reward of $\sim$6000~\AA.  We addressed this problem by using the type dG-K standard star  G158-100, whose flux peaks near 5000~\AA\ and declines rapidly towards the blue.  Flux calibration derived from this standard produced a good match between the blue- and red-side spectra of our QSOs.  The blue (red) side was extracted in wavelength bins of 120 (60) \kms.

Out of this sample, 23 QSOs were selected to create the composite described in Section~\ref{sec:composites}.  These objects are listed in Table~\ref{tab:qsos} and their spectra are plotted in Appendix~{\ref{app:spectra}}.  The remaining four QSOs were rejected either due to the presence of BAL features (J2245$+$0024, J0210$+$0003, J0218$-$002) or due to difficulty in measuring the redshift (J0215$-$0529).

\subsection{ESI and X-Shooter spectra}\label{sec:z6}

Our $z \sim 6$ sample is drawn from the Keck ESI and VLT X-Shooter spectra used by \citet{becker2019}.  A lower redshift bound of $z > 5.8$ was chosen so that the entire spectrum blueward of the \lya\ emission line down to a rest-frame wavelength of 820~\AA\ falls entirely in the VIS arm of X-Shooter.  An upper bound of $z < 6.1$ was chosen so that the Lyman series opacity of the IGM still allows some possibility of measuring flux blueward of the Lyman limit.  Due to the high sensitivity required to detect any continuum transmission at these redshifts, we also required a minimum signal-to-noise ratio in the continuum near rest-frame 1285~\AA\ of $S/N \ge 20$ per 30~\kms\ interval.  After rejecting BALs and objects with strong associated metal absorption and/or associated \lya\ damping wing absorption (typically with associated narrow metal lines), we selected 13 QSOs.  These are listed in Table~\ref{tab:qsos}.  The mean redshift in this sample is $\langle z_{\rm qso} \rangle = 5.97$.  As described in \citet{becker2019}, the ESI spectra have a typical resolution of ${\rm FWHM} \simeq 45$~\kms\ and were extracted in bins of 15~\kms, while the X-Shooter spectra have a typical resolution of ${\rm FWHM} \simeq 25$~\kms\ in the VIS arm and were extracted in bins of 10~\kms.  Individual spectra are plotted in Appendix~\ref{app:spectra}.  The rarity of obvious transmitted flux blueward of the Lyman limit highlights the challenge of directly measuring \mfp\ at these redshifts.

\subsection{QSO redshifts}\label{sec:redshifts}

Following \citet{worseck2014}, we measured QSO redshifts from the apparent start of \lya\ forest absorption, \zforest.  Five of our $z \sim 6$ objects also have precise systemic redshifts measured from either [\ion{C}{ii}] 158 $\mu$m emission or narrow nebular \lya\ emission (see references listed in Table~\ref{tab:qsos}).  An additional six QSOs\footnote{The additional QSOs are CFHQS J2100$-$1715, PSO J065$-$26, PSO J359$-$06, ULAS J1319$+$0950, and VIK J2318$-$3029, for which we use CO redshifts from \citet{venemans2020}, and CFHQS J1509$-$1749, for which we use the CO redshift from \citet{decarli2018}.} from \citet{becker2019} have CO redshifts but were not included in the composite because they were at slightly higher redshifts or their spectra fell below our S/N requirement.  We used the combined sample of eleven objects to estimate the error in our \zforest\ estimates, finding that \zforest\ was lower than the systemic redshift by an average of 180~\kms, with a standard deviation of 180~\kms.  For LRIS, ESI, and X-Shooter QSOs without a systemic redshift measurement we offset the \zforest\ measurements by this amount to arrive at an adopted systemic redshift.  The results are listed in Table~\ref{tab:qsos}.  Given the decrease in the opacity of the \lya\ forest from $z \sim 6$ to 5 and the somewhat lower resolution of the red-side LRIS spectra versus the X-Shooter and ESI spectra, it is not entirely clear that the same offset should apply to our \zforest\ estimates at $z \sim 5$.  On the other hand, 180~\kms\ corresponds to an offset 0.32~pMpc at $z = 5$, which is relatively small compared to the statistical uncertainties in our measurement of \mfp\ at that redshift arising from cosmic variance \citep[see also][]{worseck2014}.  We therefore adopt this correction to the \zforest\ measurements for the LRIS spectra.  Redshifts for the GGG sample are adopted from \citet{worseck2014}.

\subsection{Composite spectra}\label{sec:composites}

\begin{figure}
   \begin{center}
   \includegraphics[width=0.45\textwidth]{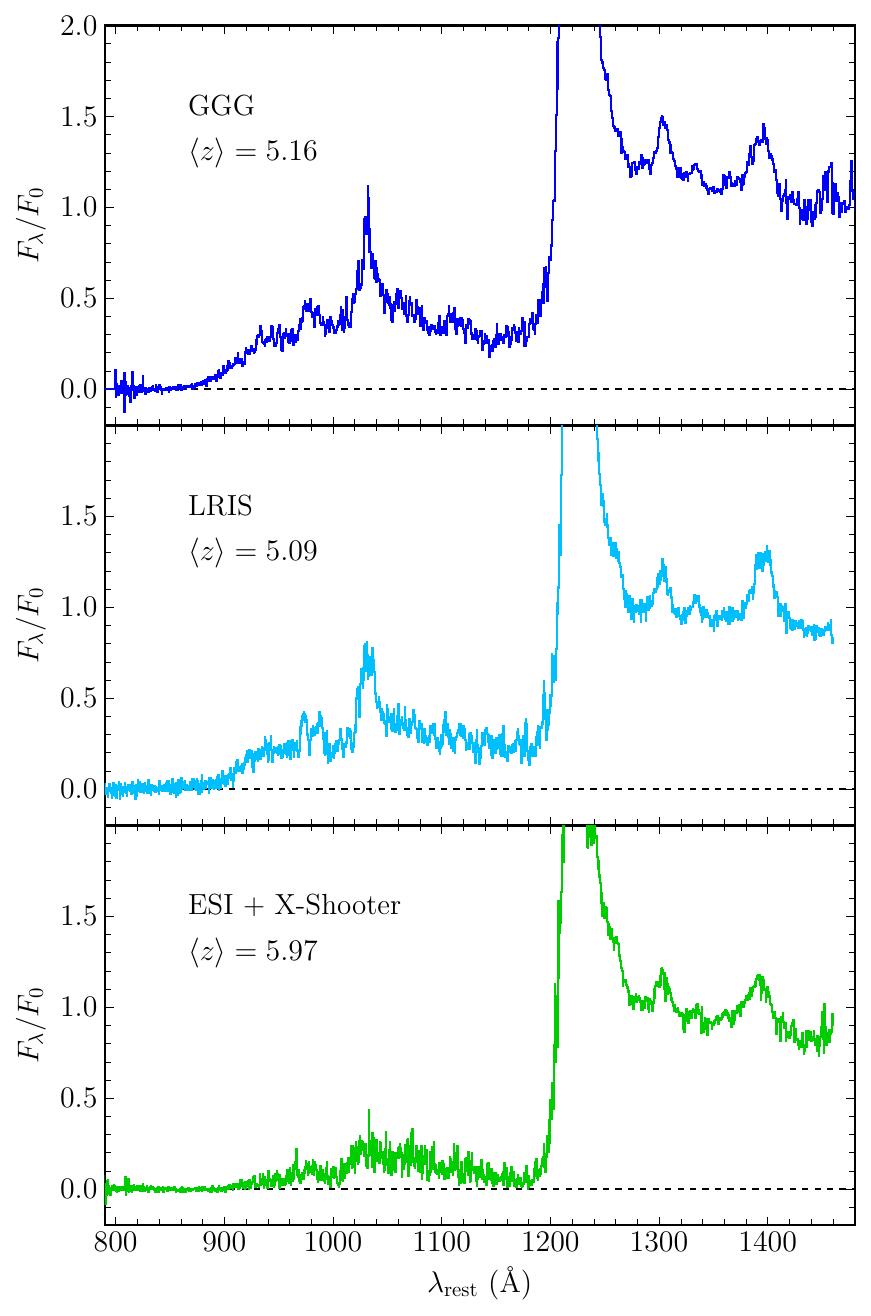}
   \vspace{-0.05in}
   \caption{Composite spectra analyzed in this work.  Panels are labeled with the sample name and the mean redshift of the QSOs included in the sample.  The flux per unit wavelength has been normalized by the continuum flux near rest-frame 1450~\AA\ for the GGG sample and over 1270--1380~\AA\ for the LRIS and ESI + X-Shooter samples.  Details of the Lyman continuum flux profiles are shown in Fig~\ref{fig:fits}.}
   \label{fig:composites}
   \end{center}
\end{figure}

We created composite spectra from each of our three samples using the following procedure.  We first shifted each spectrum to rest-frame wavelengths.  We then divided each spectrum by its continuum flux measured over wavelengths where the flux from broad emission lines is minimal.  For the GGG spectra we used the continuum flux near 1450~\AA, following \citet{worseck2014}, while for LRIS, ESI, and X-Shooter we used the median flux over 1270--1380 \AA.  The choice of wavelength range for the continuum estimate has little impact on results because the normalization of the Lyman continuum profile is treated as a free parameter.  For the LRIS, ESI, and X-Shooter spectra we corrected for residual zero-point errors by subtracting the median flux measured over a wavelength range expected to be free of transmitted flux.  These wavelength ranges (750--800~\AA\ rest frame for LRIS and 820--860 \AA\ for ESI and X-Shooter) were verified to lie well blueward of where the fitted profiles reach zero flux (see Section~\ref{sec:fits}).  For the $z \sim 6$ data the lower wavelength bound was chosen to avoid the noisy edge of the X-Shooter VIS coverage, as well residuals from the 5577 \AA\ skyline.  The zero-point estimates for these spectra were subtracted prior to creating the composites; however, we do not require the corrections to be perfect.  For the GGG sample, moreover, the wavelength coverage of the blue-side spectra does not provide a window where the zero point can be estimated safely blueward of the edge of the transmitted flux.  In all cases, therefore, we include the zero point as a free parameter when fitting models to a composite.  For an alternate treatment of the zero-point errors in the GGG data see \citet{worseck2014}.  

Wavelength regions affected by skyline subtraction residuals were identified via peaks in the error arrays and masked.  The ESI and X-Shooter spectra were also lightly median filtered using a 3-pixel sliding window to reject spurious bad pixels.  Mean composite spectra were then computed in bins of 170~\kms\ for GGG (similar to the binning used by \citealt{worseck2014}) and 120~\kms\ for the LRIS and ESI + X-Shooter data.  The results are shown in Fig.~\ref{fig:composites}.

In the measurements described below we use bootstrap resampling to estimate the uncertainty in \mfp.  In each realization, we randomly select $n_{\rm qso}$ objects from each sample, with replacement, where $n_{\rm qso}$ corresponds to the numbers in Table~\ref{tab:samples}.  Before creating the new composite we add a random redshift offset to each spectrum (excluding those with [\cii] 158 $\mu$m or nebular \lya\ redshifts) drawn from a Gaussian distribution with $\sigma = 180$~\kms (see Section~\ref{sec:redshifts}).  As noted by \citet{worseck2014}, we found that the redshift errors produce an uncertainty in \mfp\ that is small compared to the uncertainty from cosmic variance.  We nevertheless include them for completeness.  The bootstrap trials are also used to estimate the pixel-to-pixel errors in the flux, which we smooth using a polynomial fit over the wavelength range used to measure \mfp.  Additional sources of error are described in Section~\ref{sec:Req}.

When fitting the composites we include wavelengths down to 826~\AA\ for GGG, which is limited by the wavelength coverage of the data.  For LRIS we fit down to 800~\AA, while for ESI + X-Shooter we fit down to 820~\AA.  We note that wavelength range used to fit the composite overlaps with the wavelength range used to measure zero point offsets in the ESI and X-Shooter spectra.  We find, however, that this choice does not have a significant impact on our results.  The upper bound in wavelength is 910~\AA\ for all composites, a choice we describe in Section~\ref{sec:formalism}.

\section{Mean Free Path Measurements}\label{sec:measurements}

\subsection{Formalism}\label{sec:formalism}

We measure a mean free path from the composite spectra using an approach based on the method first developed by \citet{prochaska2009} and adapted by \citet{worseck2014} to higher redshifts.  The major change included here is to allow the ionizing opacity of the IGM to scale with the local photoionization rate.  As demonstrated below, this change is necessary for extending the direct measurement method to $z \sim 6$.

The observed flux, $f_{\lambda}^{\rm obs}$, will be the mean intrinsic QSO spectral energy distribution, $f_{\lambda}^{\rm SED}$, attenuated by the effective Lyman series opacity of the foreground IGM, $\tau_{\rm eff}^{\rm Lyman}$, and the Lyman continuum effective optical depth, $\tau_{\rm eff}^{\rm LyC}$,
\begin{equation}\label{eq:fobs}
f_{\lambda}^{\rm obs} = f_{\lambda}^{\rm SED} \exp{\left( -\tau_{\rm eff}^{\rm Lyman} \right)} \exp{\left( -\tau_{\rm eff}^{\rm LyC} \right)} + f_{0}\, .
\end{equation}
Here, $f_{0}$ is a zero-point correction that we include as a free parameter (see Section~\ref{sec:composites}).  We discuss the foreground Lyman series opacity in Section~\ref{sec:lyman_series}.  The intrinsic SED blueward of the Lyman limit is modeled as a power law of the form $f_{\lambda}^{\rm SED} = f_{912} \left( \frac{\lambda}{912~\mbox{\small \AA}} \right)^{-\alpha_{\lambda}^{\rm ion}}$.  The normalization $f_{912}$ is treated as a free parameter that incorporates the intrinsic QSO SED, any Lyman continuum attenuation directly associated with the QSOs, and any relative flux calibration error between 912~\AA\footnote{Throughout this paper we use 912~\AA\ to represent the Lyman limit wavelength of 911.76 \AA.} and the rest-frame wavelengths at which the individual QSO spectra are normalized.  We adopt a nominal power-law exponent of $\alpha_{\lambda}^{\rm ion} = 0.5$ (see Section~\ref{sec:Req}).  Given how rapidly $\tau_{\rm eff}^{\rm Lyman}$ and $\tau_{\rm eff}^{\rm LyC}$ evolve with wavelength, we find that our results for \mfp\ are highly insensitive to this choice except as it impacts our calculations for the ionizing luminosity of a QSO (see Section~\ref{sec:Req}).

The effective Lyman continuum opacity for a photon emitted at redshift $z_{\rm qso}$ that redshifts to 912~\AA\ at redshift $z_{912}$ will be
\begin{equation}\label{eq:taueff_LyC}
\tau_{\rm eff}^{\rm LyC} (z_{912}, z_{\rm qso}) = \frac{c}{H_0 \Omega_{\rm m}^{1/2}} \left(1+ z_{912} \right)^{2.75} \int_{z_{912}} ^{z_{\rm qso}} \kappa_{912}(z') \left( 1+z' \right)^{-5.25} dz' \, ,
\end{equation}
where $\kappa_{912}(z)$ is the Lyman continuum opacity at 912~\AA\ at redshift $z$ \citep{prochaska2009}.  The wavelength dependence of the ionizing absorption cross-section is approximated here as $\sigma(\lambda) \propto \lambda^{-2.75}$ following \citet{omeara2013} and \citet{worseck2014}.  

Previous works at $z \ge 3$ have held $\kappa_{912}$ fixed when fitting a single QSO composite spectrum \citep{prochaska2009,fumagalli2013,worseck2014}.  The difficulty with this approach at $z > 5$, however, is that \mfp\ may become comparable to or smaller than a typical QSO proximity zone.  If the ionizing flux from a QSO decreases the opacity of the IGM in its proximity zone then this will lead to a measurement of \mfp\ that is biased high with respect to its value far from the QSO \citep[see discussions in][]{worseck2014,daloisio2018}.  This effect can be diminished by selecting QSOs that are relatively faint and hence have shorter proximity zones, as we have done for the LRIS sample.  The measurement may still be biased, however, depending on the intrinsic value of \mfp.  At $z \sim 6$, moreover, \mfp\ is expected to be significantly shorter than the typical proximity zone of any QSO bright enough to obtain a useful spectrum. 

We therefore attempt to account for the proximity effect by modeling the impact of ionizing flux from a QSO on the Lyman continuum attenuation in its vicinity.  We parametrize the dependence of the opacity on the local \hi\ ionization rate, $\Gamma$, as a power law of the form
\begin{equation}\label{eq:kappa}
\kappa_{912} = \kappa_{912}^{\rm bg} \left( \frac{\Gamma}{\Gamma_{\rm bg} }\right)^{-\xi} \, ,
\end{equation}
where $\kappa^{\rm bg}_{912}$ is the background opacity and $\Gamma_{\rm bg}$ is the average background photoionization rate.\footnote{Here, "background" quantities refer to spatially averaged values in the absence of the QSO.  We test the case where fluctuations in the UV background are present in Section~\ref{sec:mocks}.}    This form is motivated by analytic models of the IGM opacity \citep{miralda-escude2000, furlanetto2005}, as well as radiative transfer simulations of Lyman limit systems \citep{mcquinn2011a}.  These studies suggest values of $\xi \sim 2/3$ at $z>5$, which has been adopted in recent models of the \lya\ forest opacity fluctuations at these redshifts \citep{davies2016, daloisio2018, nasir2020}. The uniform opacity model used by \citet{worseck2014} corresponds to $\xi = 0$.  We discuss our priors on $\xi$ further in Section~\ref{sec:xi}.

\begin{figure*}
   \centering
   \begin{minipage}{\textwidth}
   \begin{center}
   \includegraphics[width=1.0\textwidth]{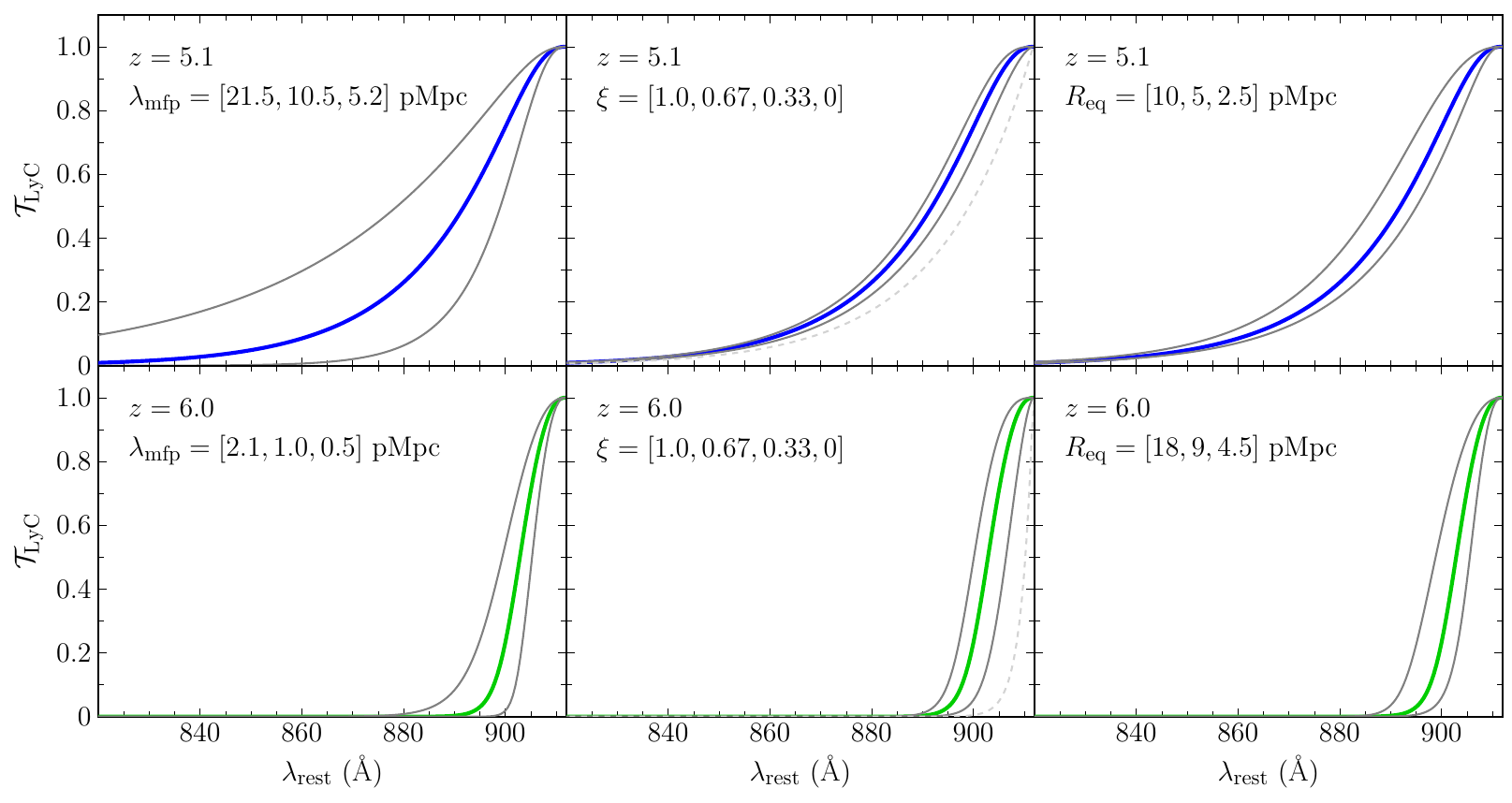}
   \vspace{-0.05in}
   \caption{Examples of Lyman continuum transmission for different model parameters.  At $z = 5.1$ (top row), the thick lines show the fiducial model with $\lambda_{\rm mfp} = 10.5$~pMpc, $\xi = 0.67$, and $R_{\rm eq} = 5$~pMpc.  At $z = 6.0$ (bottom row), the thick lines show the fiducial model with $\lambda_{\rm mfp} = 1.0$~pMpc, $\xi = 0.67$, and $R_{\rm eq} = 9$~pMpc.  The left, center, and right columns demonstrate how the transmission changes with \mfp, $\xi$, and \Req, respectively.  The left-to-right ordering of the parameters listed in brackets corresponds to the left-to-right ordering of the lines in each panel.  In the center column, the dashed line shows the $\xi = 0$ case.  Note that these transmission profiles include only Lyman continuum opacity.}
   \label{fig:ll_profile}
   \end{center}
   \end{minipage}
 \end{figure*}

The local photoionization rate will be the sum of the background rate and the contribution from the QSO, which decreases with distance, giving $\Gamma = \Gamma_{\rm bg} + \Gamma_{\rm qso}(r)$.  The Lyman  limit opacity will therefore increase with distance from the QSO as
\begin{equation}\label{eq:kappa}
\kappa_{912}(r) = \kappa_{912}^{\rm bg} \left[ 1 + \frac{\Gamma_{\rm qso}(r)}{\Gamma_{\rm bg}} \right]^{-\xi} \, .
\end{equation}
Following \citet{calverley2011}, we characterize the ionizing luminosity of a QSO relative to the ionizing background
according to the distance from the QSO, \Req, at which $\Gamma_{\rm qso}$ would be equal to $\Gamma_{\rm bg}$ in the absence of any absorption or redshifting of ionizing photons from the QSO.  We note that the actual distance at which $\Gamma_{\rm qso} = \Gamma_{\rm bg}$ will tend to be less than  $R_{\rm eq}$ due to absorption.  Nevertheless, $R_{\rm eq}$ is a convenient parameter for helping to quantify how $\kappa_{912}$ is modified near a QSO.\footnote{We echo the discussion in \citet{calverley2011} that $R_{\rm eq}$ differs from the observational definition of proximity zone size applied elsewhere at $z \gtrsim 6$.  $R_{\rm eq}$ is calculated directly from a QSO's ionizing spectrum and $\Gamma_{\rm bg}$.  It is therefore effectively a prediction for the distance to which the ionizing flux from a QSO would dominate over the background in the absence of any attenuation.  Observationally, in contrast, the proximity zone ``size'' at $z \gtrsim 6$ is typically the distance from a QSO out to which the fraction of transmitted \lya\ flux exceeds 10\% \citep[e.g.,][]{fan2006a,carilli2010}, and is therefore a measure of where the total (QSO + background) ionization rate drops below the level required for the IGM to meet this transmission threshold.  A $z=6$ QSO with $M_{1450} = -27.0$ would have $R_{\rm eq} = 11.4$ pMpc for the nominal parameters given in Section~\ref{sec:Req}.  This is roughly twice the typical proximity zone size measured by \citet{eilers2017} for QSOs near this luminosity.  This suggests, perhaps not surprisingly, that at $z \gtrsim 6$ the ionizing flux from the QSO can dominate over the background out to distances that are significantly larger than those indicated by the extent of  the observed \lya\ transmission.}  For a QSO with luminosity $L_{1450}$ at rest-frame 1450~\AA\ and a broken power-law continuum of the form
\begin{equation}\label{eq:sed}
L_{\nu}(\nu) \propto \begin{cases}
\nu^{-\alpha_{\nu}^{\rm UV}}, & 912~\mbox{\AA} < \lambda < 1450~\mbox{\AA} \\
\nu^{-\alpha_{\nu}^{\rm ion}}, & \lambda < 912~\mbox{\AA}
\end{cases}
\end{equation}
the luminosity at 912~\AA\ will be $L_{912} = L_{1450} (\nu_{912}/\nu_{1450})^{-\alpha_{\nu}^{\rm UV}}$ and this distance will be
\begin{equation}\label{eq:Req}
R_{\rm eq} = \left[ \frac{L_{912} \, \sigma_{0}}{4 \pi \, \Gamma_{\rm bg} \, (\alpha_{\nu}^{\rm ion} + 2.75)} \right]^{1/2} \, .
\end{equation}
Here, $\sigma_{0}$ is the \hi\ ionization cross-section at 912~\AA.  We calculate $L_{1450}$ from the absolute magnitudes listed in Table~\ref{tab:qsos}.  In Section~\ref{sec:Req} we calculate mean $R_{\rm eq}$ values for our samples and discuss constraints on $\Gamma_{\rm bg}$, $\alpha_{\nu}^{\rm UV}$, and $\alpha_{\nu}^{\rm ion}$.  The ionizing flux from the QSO will be diluted geometrically and attenuated by Lyman continuum absorption, which increases with distance as $\Gamma_{\rm qso}$ decreases.  We therefore solve for $\Gamma_{\rm qso}(r)$ and $\kappa_{912}(r)$ numerically under the assumption that $\kappa_{912}(r=0) = 0$.  Specifically, we divide the line of sight into small steps of distance $\delta r$.  For the first step we assume that $\Gamma_{\rm qso}$ decreases purely geometrically, i.e.,
\begin{equation}\label{eq:Gamma_qso_0}
    \Gamma_{\rm qso}(r=\delta r) = \Gamma_{\rm bg} \left( \frac{\delta r}{R_{\rm eq}} \right)^{-2} \, .
\end{equation}
Over subsequent steps we solve for $\Gamma_{\rm qso}(r + \delta r)$ as 
\begin{equation}\label{eq:Gamma_qso_r}
    \Gamma_{\rm qso}(r + \delta r) = \Gamma_{\rm qso}(r) \left( \frac{r + \delta r}{r} \right)^{-2} e^{-\kappa_{912}(r) \delta r} \, ,
\end{equation}
where $\kappa_{912}(r)$ is computed using equation~(\ref{eq:kappa}).

In principle, $L_{912}$ in equation~(\ref{eq:Req}) could be modified by an escape fraction, $f_{\rm esc}$ \citep[e.g.,][]{cristiani2016}.  For simplicity, however, we assume that the QSOs in our sample are roughly bimodal in terms of their escape fraction, having either $f_{\rm esc} \sim 0$ or 1, with $f_{\rm esc}$ independent of luminosity.  Other than cases where there is an obvious, strong associated absorber such as DLA (see Section~\ref{sec:z6}), we do not wish to bias our results by attempting to exclude QSOs with low $f_{\rm esc}$.   Fortunately, QSOs with $f_{\rm esc} = 0$ will have zero flux blueward of 912~\AA.  Including these objects should therefore only rescale the mean Lyman continuum profile, which will be captured by the normalization parameter, $f_{912}$.  Redshift errors may cause absorption from associated high-order Lyman series lines to be blended into the composite flux below 912~\AA.  We mitigate this by restricting our fits to $\lambda_{\rm rest} < 910$~\AA, i.e., $\sim$600~\kms\ blueward of the nominal QSO redshifts.  

In total, therefore, our model for the Lyman continuum flux includes five parameters, $f_{912}$, $f_0$, $\kappa_{912}^{\rm bg}$, \Req, and $\xi$.  The quantity we wish to obtain is the background mean free path that would be expected in the absence of the proximity effect.  The mean free path is defined here to be the distance travelled by photons (emitted at a wavelength somewhat shorter than 912 \AA) that would be attenuated by a factor of $1/e$ by Lyman continuum absorption.  In order to calculate this quantity with the proximity effect removed, we recompute the effective Lyman continuum opacity by setting $\kappa_{912} = \kappa_{912}^{\rm bg}$ in equation~(\ref{eq:taueff_LyC}).  Given the relatively short mean free path at these redshifts, we neglect any redshift evolution of $\kappa_{912}^{\rm bg}$.  We then compute \mfp\ as the distance between $\langle z_{\rm qso} \rangle$ and $z_{912}$ at which $\tau_{\rm eff}^{\rm LyC}(z_{912}, \langle z_{\rm qso} \rangle) = 1$.  

Examples of our model Lyman continuum transmission ($\mathcal{T}_{\rm LyC} = e^{-\tau_{\rm eff}^{\rm LyC}}$) are shown in Fig.~\ref{fig:ll_profile}.  The fiducial models at $z = 5.1$ and 6.0 use $[\lambda_{\rm mfp}, \xi, R_{\rm eq}] = [10.5~{\rm pMpc}, 0.67, 5.0~{\rm pMpc}]$ and $[1.0~{\rm pMpc}, 0.67, 9.0~{\rm pMpc}]$, respectively.  The \mfp\ values correspond to $\log{(\kappa^{\rm bg}_{912}/{\rm cm^{-2}})} = -25.5$ ($-$24.5) at $z = 5.1$ (6.0).  For all models we fix $f_{912} = 1$ and $f_{0} = 0$.  The fiducial models were chosen to be similar to those measured from the data (see Sections~\ref{sec:Req} and ~\ref{sec:fits}).  We then show how the profile varies with \mfp, $\xi$, and \Req.  Changes in $\xi$ and \Req\ have a wavelength (radial) dependence that is significantly different from \mfp\ because $\xi$ and \Req\ mainly impact the transmission profile within the proximity zone.  As expected, the relative importance of the proximity effect is larger at $z = 6.0$, where a change of $\pm 1/3$ in $\xi$ or a factor of two change in \Req\ produces a comparable change in the transmission profile as a factor of two change in \mfp.  Even so, these examples suggest that it is possible to measure \mfp\ at $z = 6$ given reasonable constraints on $\xi$ and \Req, even when \Req\ is a factor of ten larger than \mfp.  Our constraints on $\xi$ and \Req\ are discussed further below.

\subsection{Tests with mock spectra}\label{sec:mocks}

\begin{figure*}
   \centering
   \begin{minipage}{\textwidth}
   \begin{center}
   \includegraphics[width=1.0\textwidth]{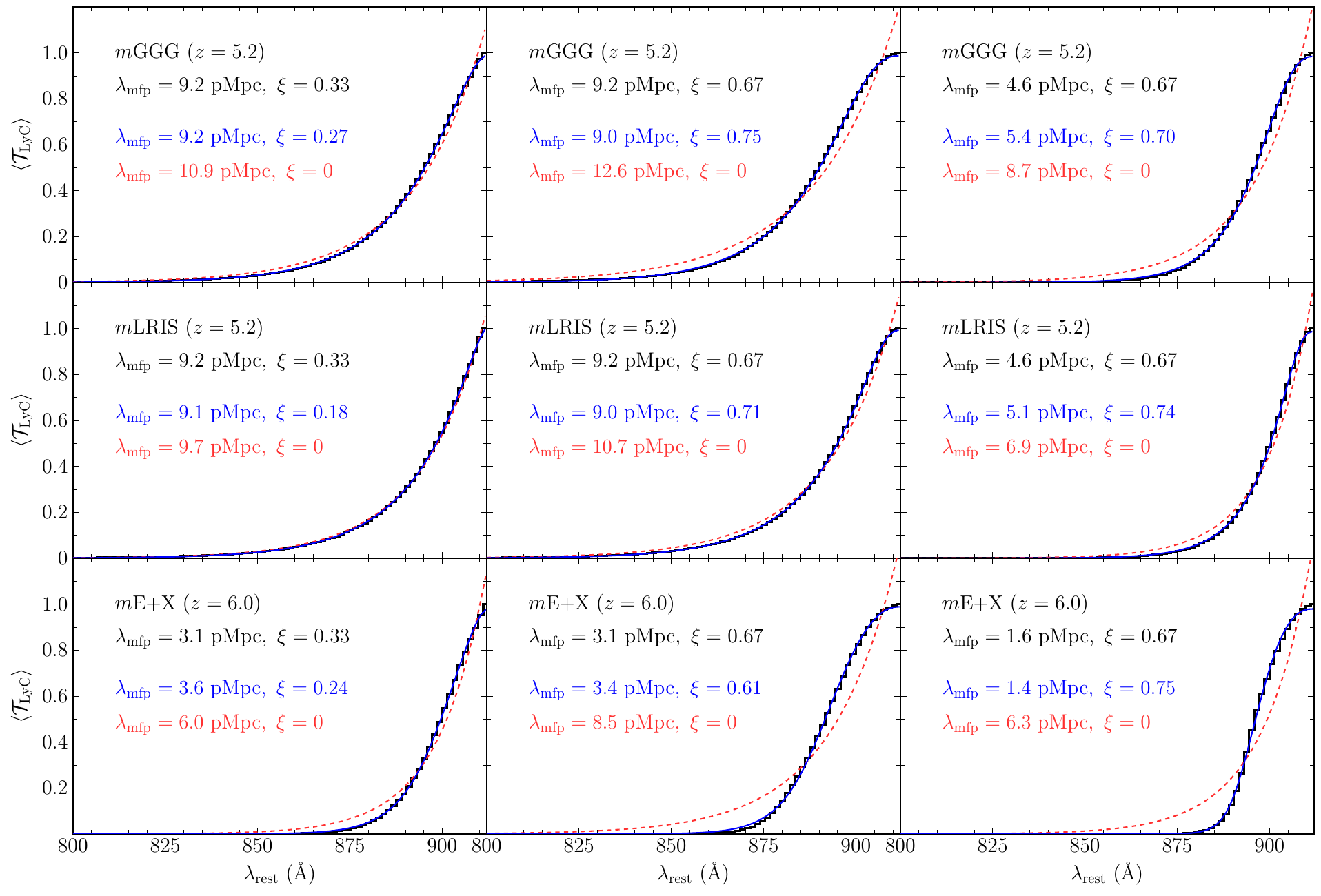}
   \vspace{-0.05in}
   \caption{Fits to mock composite Lyman continuum transmission profiles.  The top line in the legend in each panel gives the observed QSO sample to which the simulated QSO sample is matched in luminosity, along with the redshift.  The second line gives the ``true'' values of \mfp\ and $\xi$ for that mock.  Each mock spectrum (histogram) is an average of 1,000 lines of sight towards each QSO in the sample.  The solid line is a fit in which both $\kappa_{\rm bg}$ (and hence \mfp) are treated as free parameters, with results given in the third line of the legend.  The dotted line is a fit assuming constant opacity ($\xi = 0$), with results given in the fourth line.  See Section~\ref{sec:mocks} for details.}
   \label{fig:mock_profiles}
   \end{center}
   \end{minipage}
\end{figure*}

Here we investigate how well our analytic model recovers the relevant parameters from mock spectra drawn from simulations.   We refer the reader to Section 4 of \citet{daloisio2018} for a description of the simulations.  In summary, we assign QSOs with luminosities taken from the GGG, LRIS, and ESI + X-Shooter samples to the most massive halos in a cosmological hydrodynamics simulation with box $L=200~h^{-1}$ Mpc and $N_{\rm gas}= N_{\rm dm} = 2048^3$ gas and dark matter resolution elements.  The hydrodynamics simulation was run with a modified version of the code of \citet{trac2004}. The QSO halos masses range from 1.3 to $8.0 \times 10^{12}~h^{-1}\,{\rm M_\odot}$.  The QSO luminosities at 1450~\AA\ rest-frame are taken from Table~\ref{tab:qsos}.  We compute the ionizing luminosity of each QSO assuming a broken power-law of the form given by \citet{lusso2015}, which is similar to what we assume for the data (see Section~\ref{sec:Req}).

One QSO is populated in the box at a time and we use the attenuation model of \citet{davies2016} to compute the $\Gamma$ and \mfp\ fields in the box.  These iterative calculations include galactic sources, spatially varying \mfp\, and the backreaction of $\Gamma$ on local \mfp\ values. The background ionization rates are $\Gamma_{\rm bg} = 5 \times 10^{-13}$~s$^{-1}$ and $1 \times 10^{-13}$~s$^{-1}$ at $z = 5.2$ and 6, respectively, which are somewhat different than the values we use when fitting the data (see below). For the $\Gamma$ and \mfp\ computations we use uniform grids with $64^3$ cells.   We compute $1,000$ transmission profiles along random sight lines emanating from each QSO in a given sample.  We then construct $1,000$ mock composite spectra by averaging over the QSOs in the sample.  

Mock transmission profiles are generated for different combinations of \mfp\ and $\xi$ at $z = 5.2$ and 6.0.  At $z=5.2$ we consider a ``long'' mean free path model with $\lambda_{\rm mfp} = 9.2$ pMpc (40 comoving \hinvMpc) with either $\xi = 0.33$ or $0.67$, and a ``short'' mean free path model with $\lambda_{\rm mfp} = 4.6$ pMpc (20 comoving \hinvMpc) and $\xi = 0.67$.  At $z = 6.0$ we use $\lambda_{\rm mfp} = 3.1$ pMpc (15 comoving \hinvMpc) with either $\xi = 0.33$ or $0.67$, and $\lambda_{\rm mfp} = 1.6$ pMpc (8 comoving \hinvMpc) with $\xi = 0.67$.

We test how well our fitting approach recovers the ``true" \mfp\ and $\xi$ values by fitting our model to the mock composite spectra.  We compute \Req\ for each QSO using the $\Gamma_{\rm bg}$ values quoted above.  For consistency, we use the same QSO SED that was used to compute the ionizing luminosities for the mock sample.  The mean \Req\ values are  7.4, 3.5, and 16.9 pMpc for the mock GGG, LRIS, and ESI + X-Shooter samples, respectively, which we adopt when fitting the models.  These values are somewhat  larger than the values we compute for the data (see Section~\ref{sec:Req}), mainly due to the difference in $\Gamma_{\rm bg}$.  Our fits to the mock composites have three free parameters: $\kappa^{bg}_{912}$, $\xi$, and $f_{912}$.  For comparison, we also fit a constant opacity model that ignores the QSO proximity effect ($\xi = 0$).  We employ a chi-squared approach assuming equal variance in each wavelength bin.  The mocks do not include foreground Lyman series absorption or variations due to intrinsic QSO SEDs.  They therefore allow us to determine how well the \mfp\ and $\xi$ values are recovered under ideal circumstances.

Fits to the mock transmission profiles are shown in Fig.~\ref{fig:mock_profiles}.  In each case where we include the proximity effect in the fit we recover the correct \mfp\ to within 17\%.  This is true even in the ``short'' ($\lambda_{\rm mfp} = 1.6$ pMpc) case at $z = 6$, where \Req\ is a factor of ten larger than \mfp.  We also recover the correct $\xi$ to within $\sim$0.1 in all cases except the $\lambda_{\rm mfp} = 9.2$ pMpc, $\xi = 0.33$ case with the mock LRIS composite, where the impact of the proximity effect is weakest.  In contrast, ignoring the proximity effect can produce a significant overestimate of the mean free path (and overestimates of the normalization, a fact that may be evident when fitting high-$S/N$ composites).  For $\lambda_{\rm mfp} = 4.6$~pMpc and $\xi = 0.67$, the \mfp\ values returned for the LRIS and GGG mocks are too large by factors of 1.5 and 1.9, respectively.  This suggests that accounting for the proximity effect may be necessary even for fainter QSOs, depending on the true value of \mfp.  Errors for the constant opacity model are largest at $z=6$, with \mfp\ overestimated by up by factors of two to four.

In summary, we find that reasonable estimates of \mfp\ can be obtained even when the mean free path is much shorter than the proximity zone size provided that the proximity effect is taken into account.  Fitting a constant opacity model to Lyman continuum profiles at $z > 5$, in contrast, can lead to significant overestimates of the mean free path, even for samples of relatively faint QSOs.  In principle, at least, it is also possible to recover the scaling of Lyman continuum opacity with local ionization rate.  Directly constraining $\xi$ requires extremely good data, however, a point we return to below.

An important caveat is that the simulations on which we validated our technique for simultaneously fitting \mfp\ and $\xi$ do not include dynamical effects that are especially relevant if reionization ended near $z= 6$. \citet{park2016} and \citet{daloisio2020} found that impulsive changes to the UVB (e.g. reionization or a QSO turning on suddenly) shape the density structure of the IGM over $\Delta t \sim 100$ Myr through the interplay between self-shielding and hydrodynamic response of the gas to photoheating. One implication raised by \citet{daloisio2020} is that the dependence of \mfp\ on $\Gamma$ may be more complex than can be captured with a universal power law.  The simulations also assume an infinite QSO lifetime.  If the QSOs are much younger than the $\sim 100$ Myr relaxation timescale of the optically thick absorbers that set \mfp, another distinct possibility is that the local mean free paths have not had sufficient time to respond to the enhanced UV intensities. In this case, the proximity effect would be less apparent in the measurements of \mfp.

\subsection{\Req\ values for observed QSOs}\label{sec:Req}

The \Req\ estimates for our QSOs are derived from observational constraints on the metagalactic hydrogen ionization rate and the mean SED of high-redshift QSOs.  Similar to previous works \citep[e.g.,][]{becker2013a}, we estimate $\Gamma_{\rm bg}$ based on the mean intergalactic \lya\ transmission at these redshifts.  Our nominal evolution in the mean 
\lya\ transmission, described in Section~\ref{sec:lyman_series}, corresponds to $\langle \mathcal{T}_{\rm Ly\alpha} \rangle = 0.14$ at $z = 5.1$ and $\langle \mathcal{T}_{\rm Ly\alpha} \rangle = 0.0072$ at $z = 6.0$.  These values are based on measurements made from QSO spectra well outside the proximity zone (see below).  We use a hydrodynamical simulation to translate these $\langle \mathcal{T}_{\rm Ly\alpha} \rangle$ values into $\Gamma_{\rm bg}$ estimates by rescaling the simulated UV background such that the mean \lya\ transmission of the simulation box matches observations.  Specifically, we use the 40~\hinvMpc\ box with $2 \times 2048^3$ particles (40-2048) from the Sherwood simulation suite \citep{bolton2017}, whose IGM temperatures over $5 < z < 6$ are broadly consistent with existing measurements \citep{bolton2012,boera2019,walther2019,gaikwad2020}.  This procedure yields $\Gamma_{\rm bg} \simeq 7 \times 10^{-13}$~s$^{-1}$ and $3 \times 10^{-13}$~s$^{-1}$ at $z = 5.1$ and 6.0, respectively.  The uncertainties affecting $\Gamma_{\rm bg}$, including those related to $\langle \mathcal{T}_{\rm Ly\alpha} \rangle$, the temperature-density relation, and numerical effects, are similar to those in \citet{becker2013a}.  We therefore adopt a similar overall error on our $\Gamma_{\rm bg}$ estimates, namely $\pm$0.15 dex.  For the QSO SED in equation~(\ref{eq:sed}) we adopt $\alpha_{\nu}^{\rm UV} = 0.6 \pm 0.1$ and $\alpha_{\nu}^{\rm ion} = 1.5 \pm 0.3$ ($\alpha_{\lambda}^{\rm ion} = 0.5 \pm 0.3$).  The choice of $\alpha_{\nu}^{\rm UV}$ is taken from fits to composite QSO spectra by \citet{lusso2015}, and is generally consistent with other similar works \citep{vandenberk2001,shull2012a,stevans2014}.  Here we adopt a larger error than found by \citet{lusso2015} in order to allow for greater sample variance.  Our choice of $\alpha_{\nu}^{\rm ion}$ is broadly consistent with fits to composite spectra from \citet{telfer2002}, \citet{stevans2014}, and \citet{lusso2015} (though see \citealt{scott2004}, who find a harder ionizing slope for low-redshift AGN).  For the above parameters and the $M_{1450}$ values listed in Table~\ref{tab:qsos} we calculate mean \Req\ values of 6.4, 3.0, and 11.1 pMpc for the GGG, LRIS, and ESI + X-Shooter samples, respectively.

For each bootstrap composite that is used to estimate the uncertainty in \mfp\ (see Section~\ref{sec:composites}) we randomly sample the above error distributions for $\Gamma_{\rm bg}$, $\alpha_{\nu}^{\rm UV}$, and  $\alpha_{\nu}^{\rm ion}$ and propagate these into the estimates of \Req\ for each object.  We then recompute the mean $R_{\rm eq}$ based on the objects in that bootstrap sample.  The same value of $\alpha_{\nu}^{\rm ion}$ is used to model the Lyman continuum transmission profile for a given bootstrap trial.  When fitting the GGG and LRIS profiles simultaneously, the same random realizations of $\Gamma_{\rm bg}$ and the QSO spectral indices are applied to both data sets.  For reference, the 68\% (95\%) ranges of the mean $R_{\rm eq}$ values from the bootstrap trials are 5.4--7.8 (4.5--9.3) pMpc, 2.5--3.6 (2.1--4.3) pMpc, and 10.3--15.1 (8.6--18.2) pMpc for the GGG, LRIS, and ESI + X-Shooter samples, respectively.

\subsection{Foreground Lyman series transmission}\label{sec:lyman_series}

\begin{figure*}
   \centering
   \begin{minipage}{\textwidth}
   \begin{center}
   \includegraphics[width=1.0\textwidth]{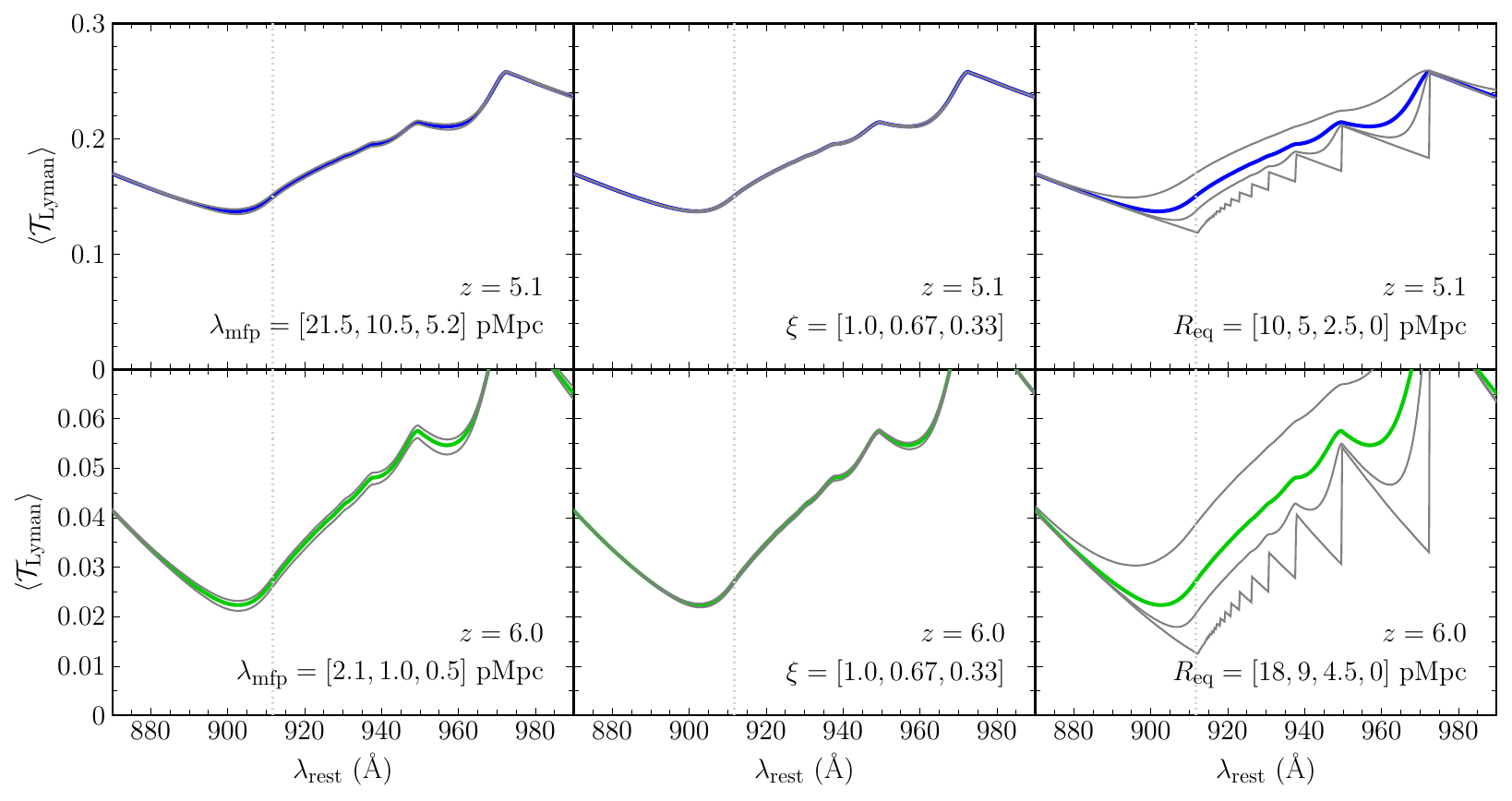}
   \vspace{-0.05in}
   \caption{Examples of Lyman series transmission for different model parameters.  At $z = 5.1$ (top row), the thick lines show the fiducial model with $\lambda_{\rm mfp} = 10.5$~pMpc, $\xi = 0.67$, and $R_{\rm eq} = 5$~pMpc.  At $z = 6.0$ (bottom row), the thick lines show the fiducial model with $\lambda_{\rm mfp} = 1.0$~pMpc, $\xi = 0.67$, and $R_{\rm eq} = 9$~pMpc.  The left, center, and right columns demonstrate how the transmission changes with \mfp, $\xi$, and \Req, respectively.  The left-to-right ordering of the parameters listed in brackets corresponds to the top-to-bottom ordering of the lines in each panel.  In the right column, the jagged line corresponds to the no proximity effect ($R_{\rm eq} = 0$) case.  Vertical dotted lines mark the Lyman limit.  Note that the model parameters only impact the change in transmission due to the proximity effect and do not self-consistently modify the baseline opacity (see Section~\ref{sec:lyman_series}).}
   \label{fig:lyman_series}
   \end{center}
   \end{minipage}
 \end{figure*}

The Lyman series opacity at an observed wavelength $\lambda_{\rm obs} < (912~\mbox{\AA})(1+z_{\rm qso})$ will include foreground contributions from all Lyman series lines,  
\begin{equation}\label{eq:taueff_lyman}
\tau_{\rm eff}^{\rm Lyman} ( \lambda_{\rm obs} ) = \sum_{j} \tau_{\rm eff}^{j} (z_j) \,
\end{equation}
where $\tau_{\rm eff}^{j} (z_j)$ is the effective opacity of transition $j$ at redshift $z_j$, $(1+z_j) \lambda_{j} = \lambda_{\rm obs}$, and $\lambda_{j}$ is the rest-frame wavelength of transition $j$. 
We compute $\tau_{\rm eff}^{\rm Lyman}$ using the 40-2048 Sherwood simulation described above.  The simulation outputs are spaced in redshift intervals of $\Delta z = 0.1$, with 5000 lines of sight drawn from each output.  At each simulation redshift we first compute baseline \lya\ optical depths by rescaling the native simulated \lya\ optical depths to reproduce the observed mean IGM \lya\ transmission.  The optical depths for 38 higher-order Lyman line are then computed as $\tau^{j} / \tau^{\alpha} = (f_j \lambda_j) / (f_{\alpha} \lambda_{\alpha})$, where $f$ here is the oscillator strength.  The mean \lya\ transmission, $\langle \mathcal{T}_{\rm Ly\alpha} \rangle = e^{-\tau_{\rm eff}^{\alpha}}$, is taken from \citet{becker2013} at $z \le 4$ and interpolated between the values of \citet{bosman2018} at  $z \ge 5.2$.  Following \citet{boera2019}, we bridge between these two sets of measurements with a power law of the form $\tau_{\rm eff}^{\alpha} (z) = 1.56 \left[ (1+z) / 5.75 \right]^{4.0}$ over $4.0 < z < 5.2$.  For reference, our adopted $\langle \mathcal{T}_{\rm Ly\alpha} \rangle$ values at $z = 3.5, 4.0$, 4.5, 5.0, 5.5 and 6.0 are 0.54, 0.41, 0.27, 0.16, 0.067, and 0.0072, respectively.  

We note that we are computing the Lyman series transmission from an optically thin simulation that does not include elements such as galactic outflows and self-shielded gas that may modify the neutral hydrogen density distribution, and hence impact the ratio of $\tau_{\rm eff}^{j} / \tau_{\alpha}$ for high-order lines.  The numerical resolution of the simulations may also have an effect.  We tested the numerical resolution using the 40-1024 run from the Sherwood suite, which also uses a 40~\hinvMpc\ box but is a factor of eight lower in mass resolution than our fiducial 40-2048 run.  Using the lower resolution run increased the total Lyman series transmission over 890$-$912~\AA\ in the rest frame by 2\% (10\%) for QSOs at $z = 5.1$ (6.0).  We tested the impact of galaxy physics using the 40-1024-ps13 runs from \citet{bolton2017}, which include a subgrid implementation of star formation and galactic outflows from \citet{puchwein2013}.  These decreased the transmission relative to the 40-1024 run by 4\% ($\sim$3\%) at $z = 5.1$ (6.0).  We also tested the impact of self-shielding using a version of the 40-1024-ps13 run in which self-shielding was added in post-processing following \citet{rahmati2013} at $z < 5$ and \citet{chardin2018a} at $z > 5$.  This decreased the mean transmission by a further 3\% (2\%) at $z = 5.1$ (6.0).  Fortunately, in all cases the effect was mainly to rescale the transmission below 912~\AA\ and not to change the shape of the profile in a way that would significantly impact our \mfp\ measurements.  These effects may nevertheless need to be considered in future works. 

An additional factor here is the QSO proximity effect.  We include the proximity effect for each Lyman series line following the same numerical approach used to compute the Lyman continuum opacities.  For a given combination of \kappabg\ and $\xi$ we compute $\tau_{\rm eff}^j$ as a function of wavelength over a grid in QSO redshift and \Req, interpolating between simulation redshifts as needed.  For each composite or bootstrap sample we then compute $\tau_{\rm eff}^{\rm Lyman} ( \lambda_{\rm obs} )$ individually for each QSO using equation~(\ref{eq:taueff_lyman}).  We then compute the transmission as $\mathcal{T}_{\rm Lyman} = \exp{\left( -\tau_{\rm eff}^{\rm Lyman} \right)}$, and average the transmission over all lines of sight.

In Fig.~\ref{fig:lyman_series} we plot the Lyman series absorption for different combinations of \mfp, $\xi$, and \Req\ at $z = 5.1$ and 6.0.  At $z = 5.1$ the transmission is not strongly affected by \mfp\ or $\xi$ because the decrease in $\Gamma_{\rm tot}$ with distance from the QSO is mainly driven by geometric dilution.  Including the proximity effect increases $\langle \mathcal{T}_{\rm Lyman} \rangle$ by a factor of $\sim$1.3 at rest-frame 912~\AA\ for $R_{\rm eq} = 5$~pMpc, similar to the mean value in the GGG sample.  It also modifies the shape of the Lyman series transmission with respect to the no proximity effect ($R_{\rm eq} = 0$) case.  At $z = 6.0$ the effect is even larger, with $\langle \mathcal{T}_{\rm Lyman} \rangle$ increasing at 912~\AA\ by a factor of 2.5 for $R_{\rm eq} = 5$ pMpc, similar to the mean value for the ESI + X-Shooter sample.  There is also a greater dependence on \mfp\ and $\xi$.  We find, however, that our final results are not highly dependent on the choice of \mfp\ and $\xi$ used for the Lyman series transmission.  When computing $\langle \mathcal{T}_{\rm Lyman} \rangle$, therefore, we hold these parameters fixed at the nominal values shown in Fig.~\ref{fig:lyman_series}, which are comparable to our best-fit results.

\subsection{Priors on $\xi$}\label{sec:xi}

The scaling of $\kappa_{912}$ with $\Gamma$ is highly uncertain, especially at the high redshifts that are relevant for this study.  From a theoretical viewpoint, the value of $\xi$ is tied to the shape of the gas density distribution function near the self-shielding threshold. Adopting the \citet{miralda-escude2000} model of IGM opacity, and assuming that the density profile of a typical self-shielding absorber is isothermal, it can be shown that $\kappa_{912}\propto \Gamma^{-2/3}$, i.e. $\xi = 2/3$ \citep{furlanetto2005, mcquinn2011a}.  Indeed, this value has been adopted in recent models of the fluctuating UVB at $z>5$ \citep[e.g.][]{davies2016,daloisio2018,nasir2020}.  Using radiative transfer simulations of self-shielding systems, \citet{mcquinn2011a} found an even steeper scaling at $z=6$ with $\xi \approx 0.75$ (see their Fig. 4 and footnote 8 of \citealt{daloisio2018}). It should be noted, however, that the radiative transfer in their study was applied in post processing to absorbers extracted from hydrodynamic simulations.  This approach misses the effect of the UVB on the density structure of the absorbers. 

More recently, \citet{daloisio2020} used fully coupled radiation hydrodynamics simulations to study self-shielding systems (see also \citealt{park2016}). Their findings suggest a more complex dependence of $\kappa_{912}$ on $\Gamma$ owing to the interplay between self-shielding and the hydrodynamic response of the gas to photoheating, which occurs on a time scale of hundreds of Myr.   We can nonetheless examine their gas density distribution functions in an attempt to gain insight into $\xi$ (see their Fig. 5).  At densities well above self-shielding, the probability distribution of $\Delta$ is reasonably approximated by $P \propto \Delta^{-1.8}$, where $\Delta$ is the gas density in units of the cosmic mean. Applying the analytic arguments of \citet{furlanetto2005} and \citet{mcquinn2011a} yields a milder scaling of $\xi \approx 0.33$.  This would be the scaling for a short time after a bright source turned on suddenly, before the gas had time to react to the impulse.  We note, however, some important caveats which suggest that $\xi$ may be larger than this. First, the $P(\Delta)$ of \citet{daloisio2020} are generally not well-described by a power law near self-shielding. Indeed, $\Delta^3 P$ appears to flatten at densities closer to self-shielding, implying a stronger dependence of $\kappa_{912}$ on $\Gamma$.  Secondly, the dependence would likely evolve as the density structure of the gas readjusted to the changing UVB. 
Based on these considerations, we argue here that $\xi = 0.33$ may serve as an approximate lower limit.  On the other hand, $\xi = 1$ is the scaling for the case of a uniform IGM in photoionization equilibrium.  This limit is approached if the opacity is dominated by diffuse gas near the mean density, rather than over-dense peaks.  In our fits we adopt a nominal value of $\xi = 0.67$ and a range $\xi = 0.33$--1.0 with a flat prior from which we randomly sample when performing bootstrap trials.  We also perform fits with $\xi$ fixed to 0.33, 0.67, and 1.0.

In principle, one can measure $\xi$ directly from the data.  Even with good constraints on \Req\ this is difficult, however, because at $z = 5.1$ the dependence of the transmitted flux on $\xi$ is relatively weak unless the mean free path is short (Fig.~\ref{fig:ll_profile}), while at $z = 6.0$ the data are too noisy to distinguish between variations in \mfp\ and $\xi$.  In a joint fit to the GGG and LRIS data we find $\xi = 0.56$, consistent with theoretical expectations, but with a 68\% (95\%) confidence range of 0.20 to 1.20 (-0.06 to 2.28).  Much of this parameter space is strongly disfavored on theoretical grounds, as described above.  The choice of $\xi$ ultimately has little impact at $z = 5.1$.  Setting $\xi = 0.33$ (1.0) increases (decreases) our nominal result by 8\% (6\%).  The impact of $\xi$ is more significant at $z = 6.0$, where the proximity effect is more pronounced.  There, setting $\xi = 0.33$ (1.0) increases (decreases) our nominal result by 69\% (68\%).  This represents a substantial portion of our error budget at $z = 6.0$.  In future works it may be possible to better constrain $\xi$ directly from the data.

\subsection{Fits to the data}\label{sec:fits}

\begin{figure}
   \begin{center}
   \includegraphics[width=0.45\textwidth]{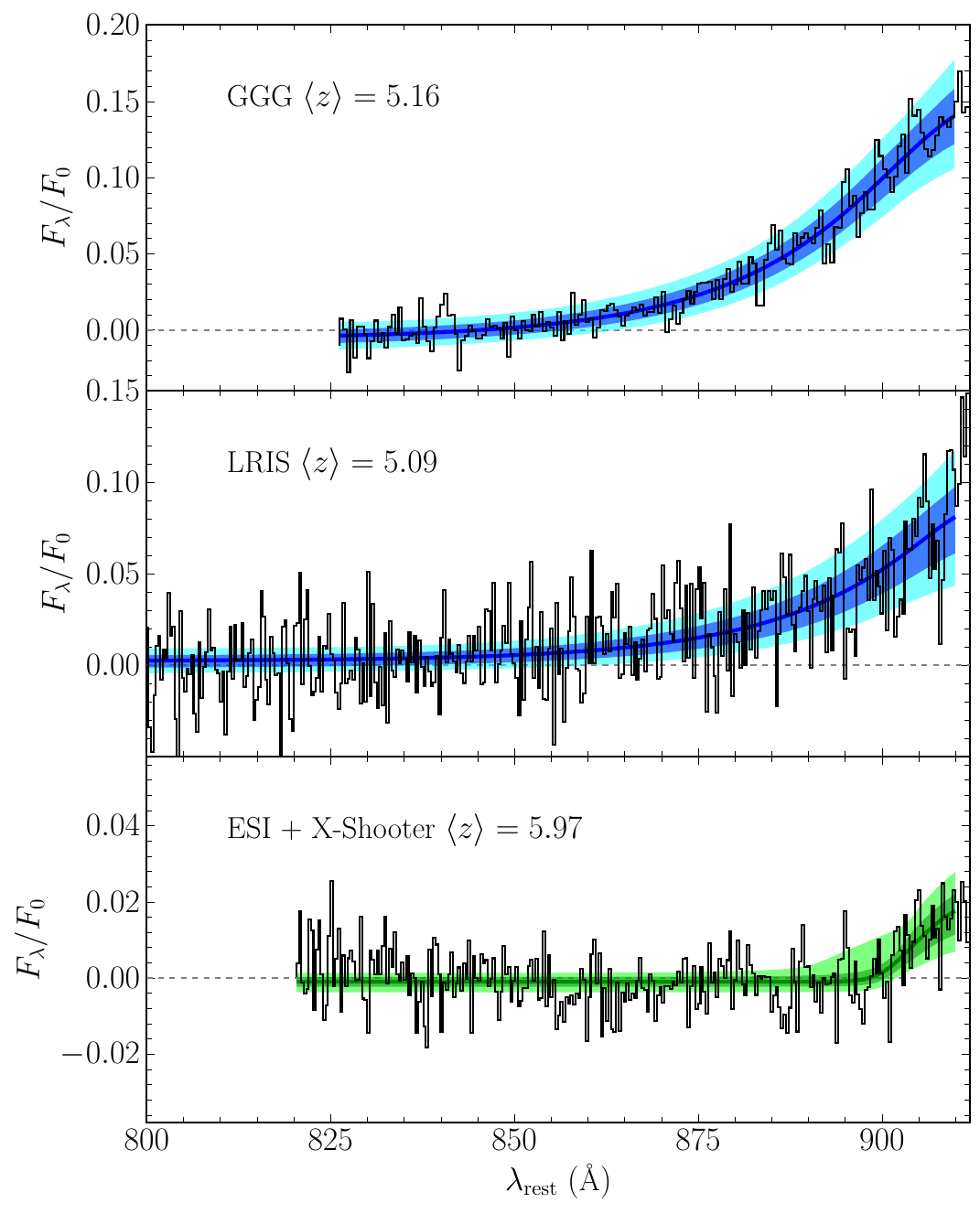}
   \vspace{-0.05in}
   \caption{Fits to the data.  The histogram in each panel shows the observed flux blueward of the Lyman limit for the sample indicated, normalized as in Fig.~\ref{fig:composites}.  Solid lines show the best-fitting model profiles, which for GGG and LRIS is a simultaneous fit to both data sets.  Dark and light shaded regions show the 68\% and 95\% intervals, respectively, spanned by fits to the bootstrap trials.}
   \label{fig:fits}
   \end{center}
\end{figure}

\begin{figure}
   \begin{center}
   \includegraphics[width=0.45\textwidth]{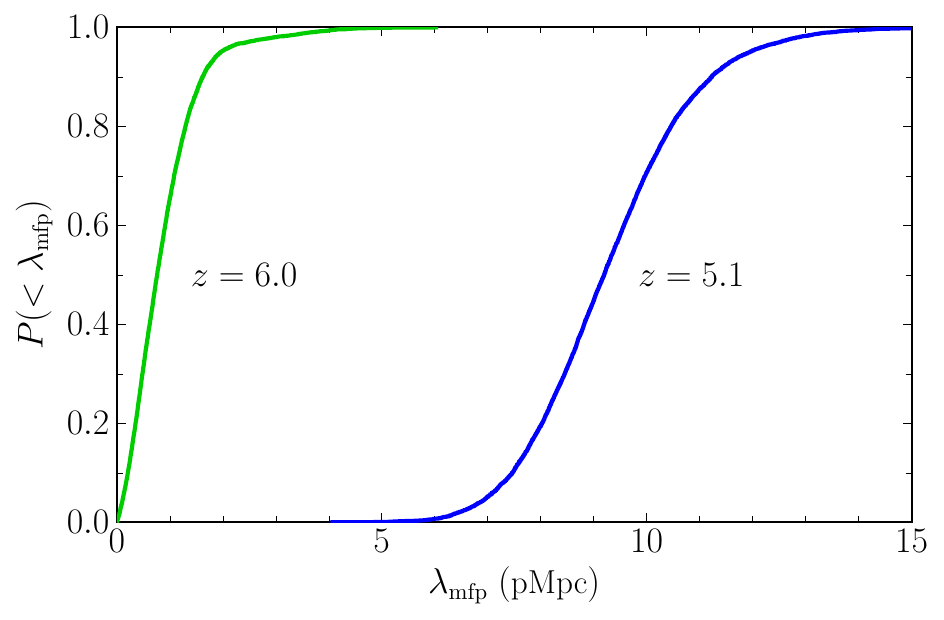}
   \vspace{-0.05in}
   \caption{Cumulative probability distribution function from bootstrap fits to \mfp\ at $z = 5.1$ (right curve) and 6.0 (left curve).}
   \label{fig:cpdfs}
   \end{center}
\end{figure}

At $z = 5.1$ we fit the GGG and LRIS composites individually as well as jointly.  For our nominal results we use $\xi = 0.67$, as noted above, and hold the mean $R_{\rm eq}$ for each composite fixed to the values given in Section~\ref{sec:Req}.  We also include the foreground Lyman series transmission described in Section~\ref{sec:lyman_series}.  We then fit for $\kappa_{912}^{\rm bg}$, which is used to calculate $\lambda_{\rm mfp}$, along with $f_{912}$ and $f_0$ separately for each composite.  In the bootstrap trials we draw $\xi$ randomly from the flat distribution over $[0.3,1.0]$ while the mean $R_{\rm eq}$ is varied according to the procedure outlined above.
From the individual fits we obtain $\lambda_{\rm mfp} = 8.85^{+1.63}_{-1.31}$~pMpc (68\% confidence intervals assuming a flat prior on $\xi$) from the GGG data and $11.64^{+4.12}_{-3.63}$~pMpc from the LRIS data.  The results are thus highly consistent with one another within the errors.  From the joint fit we obtain $\lambda_{\rm mfp} = 9.09^{+1.62}_{-1.28}$~pMpc, which we adopt as our nominal result at $z = 5.1$.  At $z = 6.0$ we measure $\lambda_{\rm mfp} = 0.75^{+0.65}_{-0.45}$~pMpc.  The nominal fits along with the ranges spanned by bootstrap trials are shown in Fig.~\ref{fig:fits}.  The cumulative probability density functions for \mfp\ at the two redshifts are shown in Fig.~\ref{fig:cpdfs}.  The main results are summarized in Table~\ref{tab:results}, where we also give results for fixed values of $\xi$.

\begin{table}
   \caption{Mean free path results}
   \label{tab:results}
   \begin{center}
   \begin{tabular*}{8.4cm}{@{\extracolsep{\fill}}lcccc}
   \hline
   $z$  &  $\xi$  &  \mfp   &  68\% range  & 95\% range \\
   \hline
   5.1  & [0.33,1.0]  &   9.09  &  7.81--10.71  &  6.58--12.66 \\
        &  0.33       &  10.06  &  8.83--11.57  &  7.78--13.37 \\
        &  0.67       &   9.09  &  7.88--10.57  &  6.77--12.32 \\
        &  1.0        &   8.34  &  7.13--9.82   &  6.00--11.55 \\
   \hline
   6.0  & [0.33,1.0]  &   0.75  &  0.30--1.40   &  0.07--2.67 \\
        &  0.33       &   1.47  &  0.91--2.02   &  0.32--3.41 \\
        &  0.67       &   0.75  &  0.37--1.36   &  0.10--2.60 \\
        &  1.0        &   0.38  &  0.17--0.93   &  0.04--2.21 \\
   \hline
   \end{tabular*}
   \begin{flushleft}
   For each redshift, the first row gives the results for \mfp\ assuming a nominal value of $\xi=0.67$ and a flat prior over $\xi = [0.33,1.0]$.  Subsequent rows give results with $\xi$ fixed.
   \end{flushleft}
   \end{center}
\end{table}

Our value of \mfp\ at $z = 5.1$ is consistent with the results from \citet{worseck2014}.  This suggests that \mfp\ at this redshift is large enough that the impact of the QSO proximity effect is relatively modest, even for the brighter GGG sample where $\langle R_{\rm eq} \rangle \simeq 0.7 \lambda_{\rm mfp}$.  Indeed, if we neglect the proximity effect by setting $\xi = 0$ and $R_{\rm eq} = 0$ for both the Lyman continuum opacity and the foreground Lyman series, emulating the approach of \citet{worseck2014}, our result for \mfp\ increases by only 12\% for the GGG composite and remains essentially unchanged for the LRIS composite.  This is somewhat less than the bias found with mock spectra in Section~\ref{sec:mocks} because the errors in the Lyman continuum modeling are partially offset by errors in the Lyman series modeling.  The difference increases to 28\% for the GGG sample if we include the proximity effect in the foreground Lyman series transmission but not in the Lyman continuum, a scenario closer to Figure~\ref{fig:ll_profile}, where accurate modeling of the Lyman series is assumed.

At $z = 6.0$, in contrast, the mean value of \Req\ is a factor of 15 larger than our value of \mfp, making it critical to take the proximity effect into account.  In this case, fully neglecting the proximity effect increases our \mfp\ measurement by a factor of 2.9 above our nominal $\xi = 0.67$ value.  If we attempt to emulate the mock trials by including the proximity effect in the Lyman series absorption but not in the Lyman continuum then our result for \mfp\ increases by a factor of 3.6.  This is consistent with the bias expected from the mock trials in Section~\ref{sec:mocks}, and emphasizes the importance of properly accounting for the proximity effect at $z \sim 6$. 

\begin{figure}
   \begin{center}
   \includegraphics[width=0.45\textwidth]{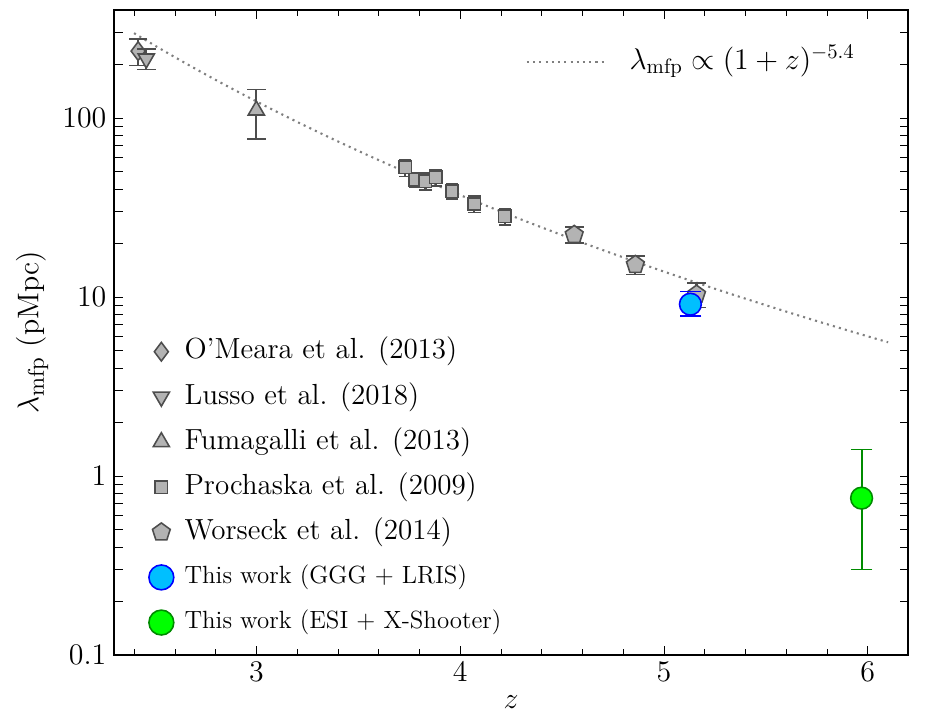}
   \vspace{-0.05in}
   \caption{Direct measurements of \mfp\ from this work and the literature.  Results from \citet{omeara2013} and \citet{lusso2018} have been offset slightly in redshift for clarity.  The dotted line shows the power-law fit to $\lambda_{\rm mfp}(z)$ over $2.44 < z < 5.16$ from \citet{worseck2014}, extrapolated out to $z=6$.}
   \label{fig:mfp}
   \end{center}
\end{figure}

In Fig.~\ref{fig:mfp} we plot our \mfp\ values as a function of redshift, along with measurements from the literature \citep{prochaska2009,fumagalli2013,omeara2013,worseck2014,lusso2018}.  The \citet{lusso2018} value at $z = 2.44$ is their fit to the data from \citet{omeara2013}.  \citet{lusso2018} find somewhat lower values of \mfp\ at $z \sim 2$ towards QSO pairs, potentially due to an increased incidence of optically thick absorbers in pair environments.  We note that \citet{romano2019} measured the mean free path towards QSOs at $z \sim 4$.  They find values that are $\sim$10--20\% higher than those of \citet{prochaska2009} and \citet{worseck2014} over the same redshifts.  In trials using the two lower-redshift GGG composites from \citet{worseck2014} we found that this discrepancy is well explained by the lack of  foreground Lyman series absorption in the \citet{romano2019} analysis.  \citet{worseck2014} fit a power law of the form $\lambda_{\rm mfp}(z) \propto (1+z)^{-5.4}$ over $2.44 < z < 5.16$ (dotted line in Fig.~\ref{fig:mfp}).  Extrapolating this fit out to $z = 6$ overshoots our nominal ESI + X-Shooter measurement by a factor of six, and is excluded by the data with $>$99.99\% confidence.  We therefore find strong evidence that the evolution of $\lambda_{\rm mfp}(z)$ with redshift steepens at $z \gtrsim 5$.  This steepening is broadly consistent with the results of \citet{songaila2010} based on their measurements of discrete Lyman limit absorbers towards QSOs over $5 < z < 6$.

\section{Discussion}\label{sec:discussion}

\subsection{Implications for reionization}

\begin{figure}
   \begin{center}
   \includegraphics[width=0.45\textwidth]{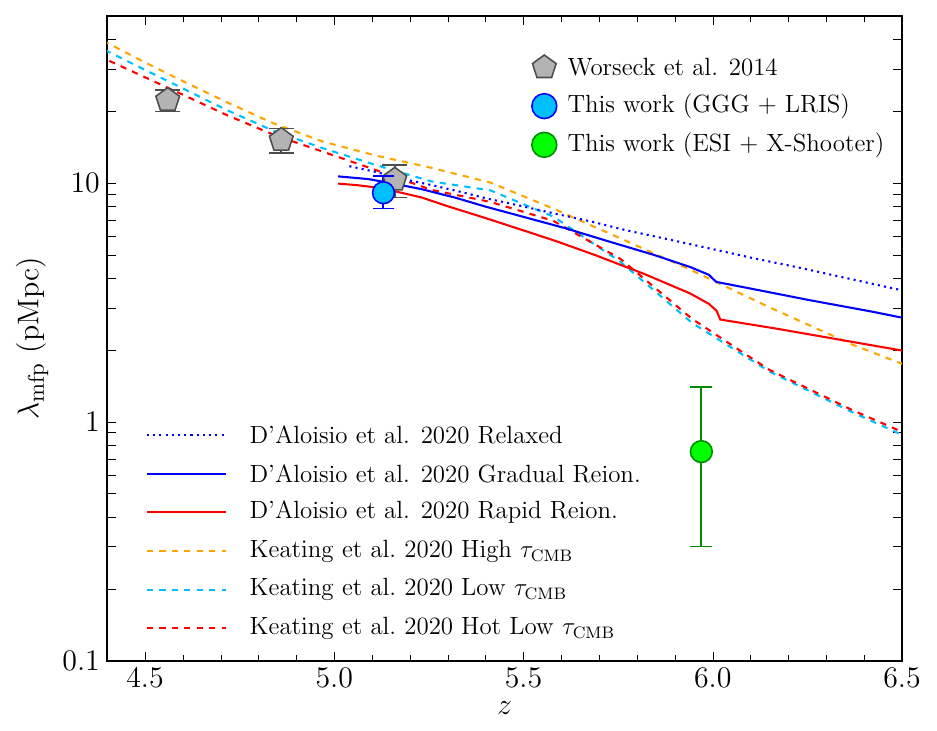}
   \vspace{-0.05in}
   \caption{Measurements of \mfp\ from this work (circles) and \citet{worseck2014} (pentagons), along with $\lambda_{\rm mfp}(z)$ relations from simulations.  Dotted and solid lines are from \citet{daloisio2020}.  The dotted line shows the expected evolution if the IGM reionized early enough that the absorbers have had time to fully relax hydrodynamically by $z = 6$.  Blue (upper) and red (lower)solid lines show their ``gradual'' and ''rapid" reionization models wherein reionization is 50\% complete at $z = 9.1$ and 7.3, respectively, and complete by $z = 6$.  Dashed lines are from \citet{keating2020a}.  The orange (upper) line shows their "High $\tau_{\rm CMB}$'' wherein reionization is 50\% complete near $z \simeq 8.4$ and ends near $z \simeq 5.3$.  The cyan (middle at $z < 5.5$) and red (lower at $z < 5.5$) dashed lines show their ``Low $\tau_{\rm CMB}$'' models wherein reionization still ends $z \simeq 5.3$ but is 50\% complete at $z \simeq 6.7$.}
   \label{fig:mfp_with_models}
   \end{center}
\end{figure}

Our measurements are consistent with a low value of \mfp\ at $z = 6$ and a rapid increase from $z = 6$ to 5.  Taken at face value, perhaps the most interesting possibility is that this evolution is tied to the end of reionization.   In Fig.~\ref{fig:mfp_with_models} we compare our measurements to predictions for \mfpz\ from different reionization models. We begin with the simplistic models in \citet{daloisio2020}, which employ results from a suite of radiative hydrodynamics simulations of the ionizing photon sinks at $z>5$. The dotted curve shows a model in which reionization ended long before $z = 6$ such that the IGM has had sufficient time to relax hydrodynamically.  This model predicts a redshift evolution of \mfp$\propto (1+z)^{-5.4}$ and a \mfp($z = 6$) that is a factor of $\sim$7 longer than our measurement.  It is worth noting that this model assumes only the $\Lambda$CDM cosmology and a constant UVB intensity; yet it yields a redshift evolution for \mfp\ that is identical to the empirical fit of \citet{worseck2014}.  The fully relaxed model is inconsistent with our $z = 6.0$ measurement at the 99.9\% level ($P(< \lambda_{\rm mfp}) = 0.999$).  For comparison, the solid curves show the ``rapid'' and ``gradual'' reionization models of \citet{daloisio2020} wherein reionization is 50\% complete at $z = 7.3$ and 9.1, respectively, and ends at $z = 6$.  Although \mfp\ at $z = 6$ is lower than in the fully relaxed models, the data are still inconsistent at the  98--99\% confidence levels.

It is also possible that reionization ended later than $z= 6$, a scenario that has been proposed recently to explain the large scatter in the $z>5$ Ly$\alpha$ forest opacity \citep{kulkarni2019,keating2020,keating2020a,nasir2020,choudhury2020,qin2021}.   The dashed curves in Fig.~\ref{fig:mfp_with_models} show the ``Low $\tau_{\rm CMB}$'' and ``Hot Low $\tau_{\rm CMB}$'' models of \citet{keating2020a}, wherein reionization is 50\% complete at $z \simeq 6.7$ and ends at $z \simeq 5.3$.  In these models the IGM at $z = 6.0$ is still $\sim$20\% neutral.  We also plot their "High $\tau_{\rm CMB}$'' model wherein reionization ends at the same redshift but is 50\% complete at $z \simeq 8.4$.  In this model the IGM at $z = 6.0$ is $\sim$8\% neutral.  The High $\tau_{\rm CMB}$ model is excluded at the 99\% level.  The Low $\tau_{\rm CMB}$ models are more consistent with our measurement at $z = 6.0$, although the data still prefer a lower \mfp\ at the 97\% confidence level.

We note that \mfp\ evolves rapidly near $z = 6$ in all of these reionization models, and that they therefore become more consistent with the data if they are shifted slightly in redshift.  For example, shifting the models by $\Delta z = -0.2$ decreases $P(< \lambda_{\rm mfp})$ at $z = 6.0$ to 0.97 for the \citet{daloisio2020} ``rapid'' model and 0.86 for the \citet{keating2020a} Low $\tau_{\rm CMB}$ model.  The low value of \mfp\ we measure at $z = 6.0$ may therefore suggest that reionization occurs even later than these models propose.

We further note that the tension with existing models may be reduced if $\xi$ is near the low end of our adopted range.  Our nominal \mfp\ value at $z = 6.0$ is a factor of two higher for $\xi = 0.33$ than for $\xi = 0.67$ (see Table~\ref{tab:results}), a result that comes from attributing less of the transmission in Figure~\ref{fig:fits} to the proximity effect.  Moreover, $P(< \lambda_{\rm mfp})$ for the Low $\tau_{\rm CMB}$ model at $z = 6.0$ decreases to 0.91 when we hold $\xi$ fixed to 0.33.  It is possible, therefore, that reconciling the reionization history with our measurements of \mfp\ may require the ionizing sinks near $z \sim 6$ to be less sensitive to photoionization effects than some models assume \citep[for further discussion, see][]{daloisio2020}.

 \subsection{Ionizing emissivity}\label{sec:emissivity} 
 
We can use our estimates of $\Gamma_{\rm bg}$ and \mfp\ to infer the ionizing emissivity at $z = 5$--6.  Here we use the local source approximation, which neglects the redshifting of ionizing photons \citep[e.g.,][]{schirber2003,kuhlen2012}.  This is a reasonable choice given the short mean free path at $z > 5$ \citep[e.g.,][]{becker2013a}.  Under this approximation the comoving ionizing emissivity is given by
\begin{equation}\label{eq:Nion}
\dot{N}_{\rm ion}(z) \approx \frac{1}{(1+z)^3} \frac{\Gamma_{\rm bg}(z)}{\sigma_{912}\lambda_{\rm mfp}(z)} \frac{(\alpha_{\rm bg} + 2.75)}{\alpha_{\rm s}} \, .
\end{equation}
Here, $\alpha_{\rm s}$ is the slope of the ionizing spectrum of the sources ($f_{\nu}^{\rm s} \propto \nu^{-\alpha_{\rm s}}$), and $\alpha_{\rm bg}$ is the slope of the ionizing background after filtering through the IGM.  If the column density distribution of \hi\ absorbers producing most of the Lyman continuun opacity is a power law of the form $f(N_{\rm H\,I}) \propto N_{\rm H\,I}^{-\beta_{N}}$, then $\alpha_{\rm bg} = \alpha_{\rm s} - 2.75(\beta_{N}-1)$.  Following \citet{becker2013a} we adopt $\alpha_{\rm s} = 2.0$ as a reasonable match to models of metal-poor star-forming galaxies, and $\beta_{N} = 1.3$.  Applying our $\Gamma_{\rm bg}$ and \mfp\ results then gives $\log{\dot{N}_{\rm ion}(z=5.1)} = 0.00^{+0.16 (0.32)}_{-0.17 (0.33)}$ and $\log{\dot{N}_{\rm ion}(z=6.0)} = 0.54^{+0.43(1.08)}_{-0.31(0.62)}$, where $\dot{N}_{\rm ion}$ is in units of $10^{51}~{\rm photons~s^{-1}~(comoving~Mpc)^{-3}}$ and the errors are 68\% (95\%) confidence intervals.  At $z = 5.1$ the errors in $\dot{N}_{\rm ion}$ are dominated by uncertainties in $\Gamma_{\rm bg}$, whereas at $z = 6.0$ they are dominated by uncertainties in \mfp.  

The results at $z = 5.1$ are similar to the values of $\dot{N}_{\rm ion}$ over $2.4 < z < 4.75$ found by \citet{becker2013a}, and suggest that the ionizing emissivity over $2 < z < 5$ may change relatively little over this period even as the source populations of star-forming galaxies and AGN evolve considerably.  In contrast, although the errors are large and we have ignored possible fluctuations in the ionizing background, the emissivity at $z = 6.0$ is potentially significantly higher.  If confirmed, this would suggest that the mean production efficiency and/or escape fraction of ionizing photons is higher for sources at $z \gtrsim 6$ than for sources at lower redshifts.  The nominal value of $\dot{N}_{\rm ion}$ at $z = 6.0$ corresponds to $\sim$17 ionizing photons per hydrogen atom per Gyr, a rate that may help explain how reionization could have been completed in only a few hundred Myr.

\subsection{Caveats and future work}

Finally, we note that this work has some limitations.  Our measurement at $z = 6.0$ is based on a relatively small sample of 13 QSOs, and within this sample there are clearly outliers in terms of Lyman continuum transmission.  The spectrum of SDSS J0836$+$0054, for example, shows discrete transmission peaks down to rest-frame 870~\AA\ (i.e., 22 pMpc from the QSO), whereas none of the other $z \sim 6$ objects shows obvious transmission below 900~\AA.  While a skewed distribution of free paths along individual lines of sight is expected \citep[e.g.,][]{romano2019}, and while this particular QSO is the brightest (and lowest-redshift) one in our ESI + X-Shooter sample, a larger sample at $z \sim 6$ would help to characterize the spatial variations in ionizing opacity near the end of reionization.  Given the rapid increase in \mfp\ between $z =6$ to 5, and the different evolutions predicted by the models over this redshift range (e.g., Figure~\ref{fig:mfp_with_models}), it also clearly of interest to constrain \mfp\ near $z = 5.5$.  

In term of the modeling, the uncertain scaling of $\kappa_{912}$ with $\Gamma$ has significant implications for \mfp\ at $z \sim 6$, as discussed above.  We also note that some of the formalism we applied herein assumes an ionized IGM.  If reionization is incomplete at $z \sim 6$, then the \mfp\ we measure at that redshift may correspond to the mean opacity only within the ionized phase provided that the ionized regions surrounding bright QSOs are larger than the proximity zone size (\Req).  The tests presented in Section~\ref{sec:mocks} suggest that our approach should be robust to the UVB fluctuations expected near the end stages of reionization.  Additional trials with more realistic late reionization simulations, however, would help to clarify how well these tools can be applied when the IGM is partly neutral.  

Finally, consistent with previous works, we have  not attempted to model the foreground Lyman series transmission in a fully self-consistent way.  Although we do not expect this to significantly impact our \mfp\ results, as discussed above, simultaneously fitting the Lyman series and Lyman continuum transmission may provide insight into the properties (e.g., the \hi\ column density distribution) of the absorbers that dominate the ionizing opacity at these redshifts.

\section{Summary}\label{sec:summary}

In this work we measure the mean free path of ionizing photons at $z \simeq 5$--6 using composite QSO spectra.  We introduce a fitting approach that accounts for the QSO proximity effect by modeling the change in ionizing opacity with the local photoionization rate.  This is also the first work to extend direct measurements of \mfp\ to $z \sim 6$, where they are sensitive to the ionizing opacity near the end of reionization.

At $z=5.1$ we measure $\lambda_{\rm mfp} = 9.09^{+1.62}_{-1.28}$~pMpc (68\% errors) from a combination of bright QSOs from the GGG survey and fainter QSOs observed with LRIS.  This is consistent with results from the GGG sample alone obtained by \citet{worseck2014}, who did not attempt to account for the proximity effect.  This suggests that \mfp\ is sufficiently long at $z \sim 5$ that the proximity effect does not greatly impact the transmission of Lyman continuum photons in QSO spectra.

At $z = 6.0$ we measure $\lambda_{\rm mfp} = 0.75^{+0.65}_{-0.45}$~pMpc using spectra from ESI and X-Shooter.  In contrast to lower redshifts, we find that neglecting the proximity effect here can bias the result high by a factor of two or more.  Our value lies well below extrapolations from lower redshifts, and suggests that the mean free path evolves rapidly over $5 < z < 6$.  A short mean free path at $z = 6.0$ and a rapid increase from $z = 6$ to 5 are qualitatively consistent with models wherein reionization ends at $z \sim 6$, or even later \citep[e.g.,][]{kulkarni2019,keating2020,keating2020a,nasir2020}, but disfavor models wherein reionization ended early enough that the IGM has had time to fully relax by $z \sim 6$ \citep[see][]{daloisio2020}.  

Models with later and more rapid reionization (i.e., the ``rapid'' model of \citealt{daloisio2020} and the ``Low $\tau_{\rm CMB}$'' models of \citealt{keating2020a}) fall closest to our \mfp\ measurements, yet our value at $z = 6.0$ lies below even models wherein the IGM at this redshift is still $\sim$20\% neutral \citep{keating2020a}.  This may indicate that the end of reionization occurred even later than previously thought.    Alternatively, the models may be missing some of the absorption systems that limit the mean free path near the end of reionization.  Further work will help to clarify how strongly the reionization history can be constrained by mean free path measurements such as the ones in this work.

\section*{Acknowledgements}

We thank Kishalay De, Anna Ho, and Yuhan Yao for assisting with the observations.  We also thank Fred Davies, Matt McQuinn, and the anonymous referee for helpful comments and suggestions.  GDB, HMC, and YZ are supported by the National Science Foundation through grant AST-1751404.  HMC is also supported by an NSF GRFP through grant DGE-1326120.  AD is supported by HST grant HST-AR15013.005-A and NASA grant 19-ATP19-0191.  JSB is supported by STFC consolidated grant ST/T000171/1. 

Some of the data presented herein were obtained at the  W.M. Keck Observatory, which is operated as a scientific partnership between the California Institute of Technology and the University of California; it was made possible by the generous support of the W.M. Keck Foundation. The authors wish to recognize and acknowledge the very significant cultural role and reverence that the summit of Maunakea has always had within the indigenous Hawaiian community. We are most fortunate to have the opportunity to conduct observations from this mountain.  This research also made use of the Keck Observatory Archive (KOA), which is operated by the W.M. Keck Observatory and the NASA Exoplanet Science Institute (NExScI), under contract with the National Aeronautics and Space Administration.  Further observations were made with ESO telescopes at the La Silla Paranal Observatory under program IDs 60.A-9024, 084.A-0390, 085.A-0299, 086.A-0162, 086.A-0574, 088.A-0897, 091.C-0934,
097.B-1070, and 098.B-0537.  This work is also based on observations obtained at the international Gemini Observatory, a program of NSF’s NOIRLab, which is managed by the Association of Universities for Research in Astronomy (AURA) under a cooperative agreement with the National Science Foundation. on behalf of the Gemini Observatory partnership: the National Science Foundation (United States), National Research Council (Canada), Agencia Nacional de Investigaci\'{o}n y Desarrollo (Chile), Ministerio de Ciencia, Tecnolog\'{i}a e Innovaci\'{o}n (Argentina), Minist\'{e}rio da Ci\^{e}ncia, Tecnologia, Inova\c{c}\~{o}es e Comunica\c{c}\~{o}es (Brazil), and Korea Astronomy and Space Science Institute (Republic of Korea).

Computational models were made possible by NSF XSEDE allocation TG-AST120066.  The Sherwood simulations were performed using the Curie supercomputer at the Tre Grand Centre de Calcul (TGCC), and the DiRAC Data Analytic system at the University of Cambridge, operated by the University of Cambridge High Performance Computing Service on behalf of the STFC DiRAC HPC Facility (www.dirac.ac.uk). This equipment was funded by BIS National E-infrastructure capital grant (ST/K001590/1), STFC capital grants ST/H008861/1 and ST/H00887X/1, and STFC DiRAC Operations grant ST/K00333X/1. DiRAC is part of the National E- Infrastructure.

\section*{Data availability}

The raw data underlying this article are available from the Gemini archive at https://archive.gemini.edu, the Keck Observatory Archive at https://www2.keck.hawaii.edu/koa/public/koa.php, and the VLT Archive at http://archive.eso.org/cms.html.   Reduced GGG data are available at the CDS via http://cdsarc.u-strasbg.fr/viz-bin/qcat?J/MNRAS/445/1745.  Other reduced data are available on reasonable request to the corresponding author.

\bibliographystyle{mnras}
\bibliography{qso_mfp_refs}

\appendix

\section{QSO Spectra}\label{app:spectra}

\begin{figure*}
   \centering
   \begin{minipage}{\textwidth}
   \begin{center}
   \includegraphics[width=1.0\textwidth]{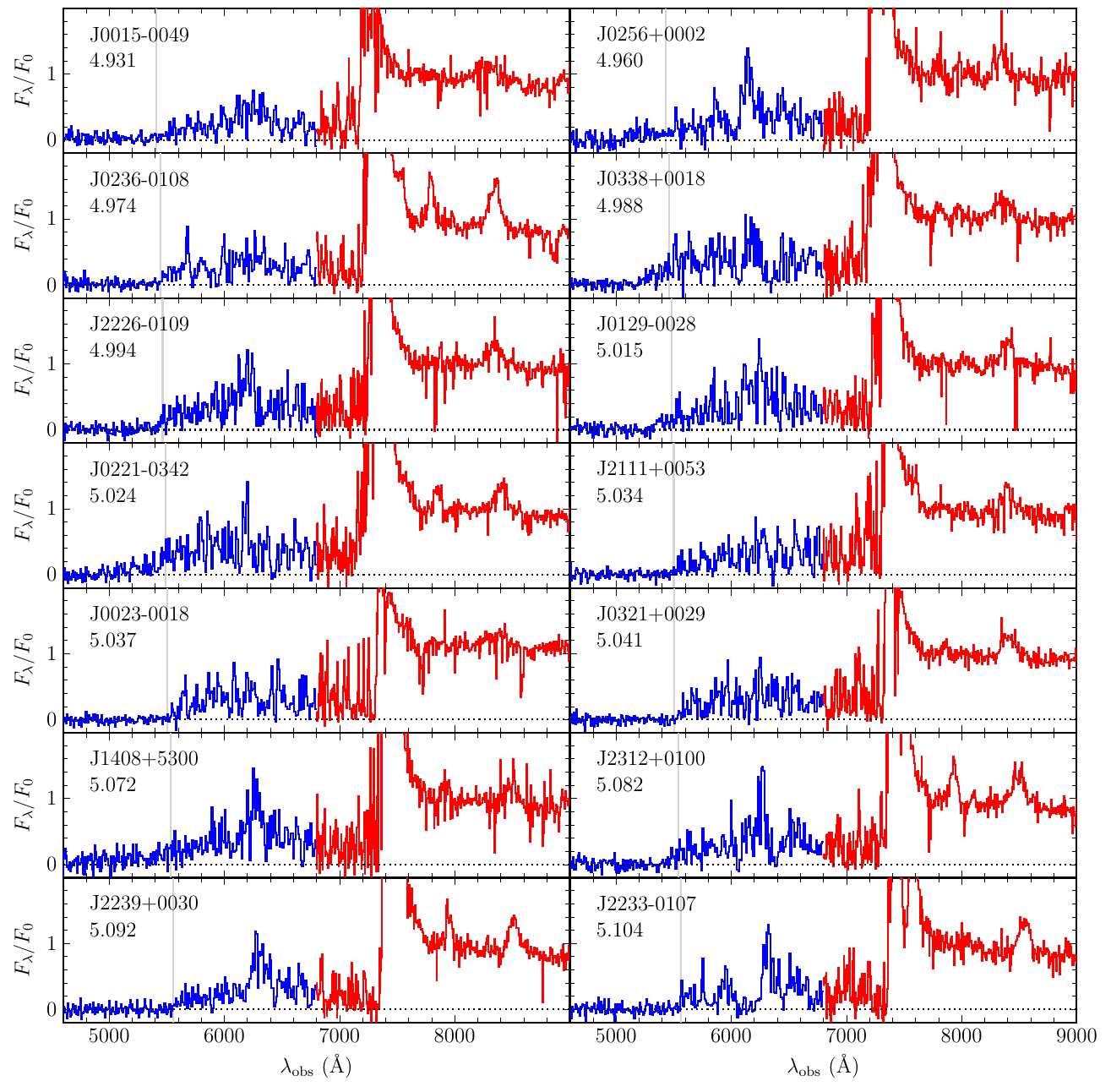}
   \vspace{-0.05in}
   \caption{LRIS spectra used in this work.  Panels are labeled with the QSO name and redshift.  For each QSO we plot flux per unit wavelength normalized by the continuum flux measured over rest-frame 1270--1380~\AA.  The blue-side data taken with the 300/5000 grism are shown in blue.  The red-side data taken with the 831/8200 grating are shown in red. Vertical lines mark the Lyman limit wavelength in the rest frame of the QSO.}
   \label{fig:lris_spectra_1}
   \end{center}
   \end{minipage}
\end{figure*}

\begin{figure*}
   \centering
   \begin{minipage}{\textwidth}
   \begin{center}
   \includegraphics[width=1.0\textwidth]{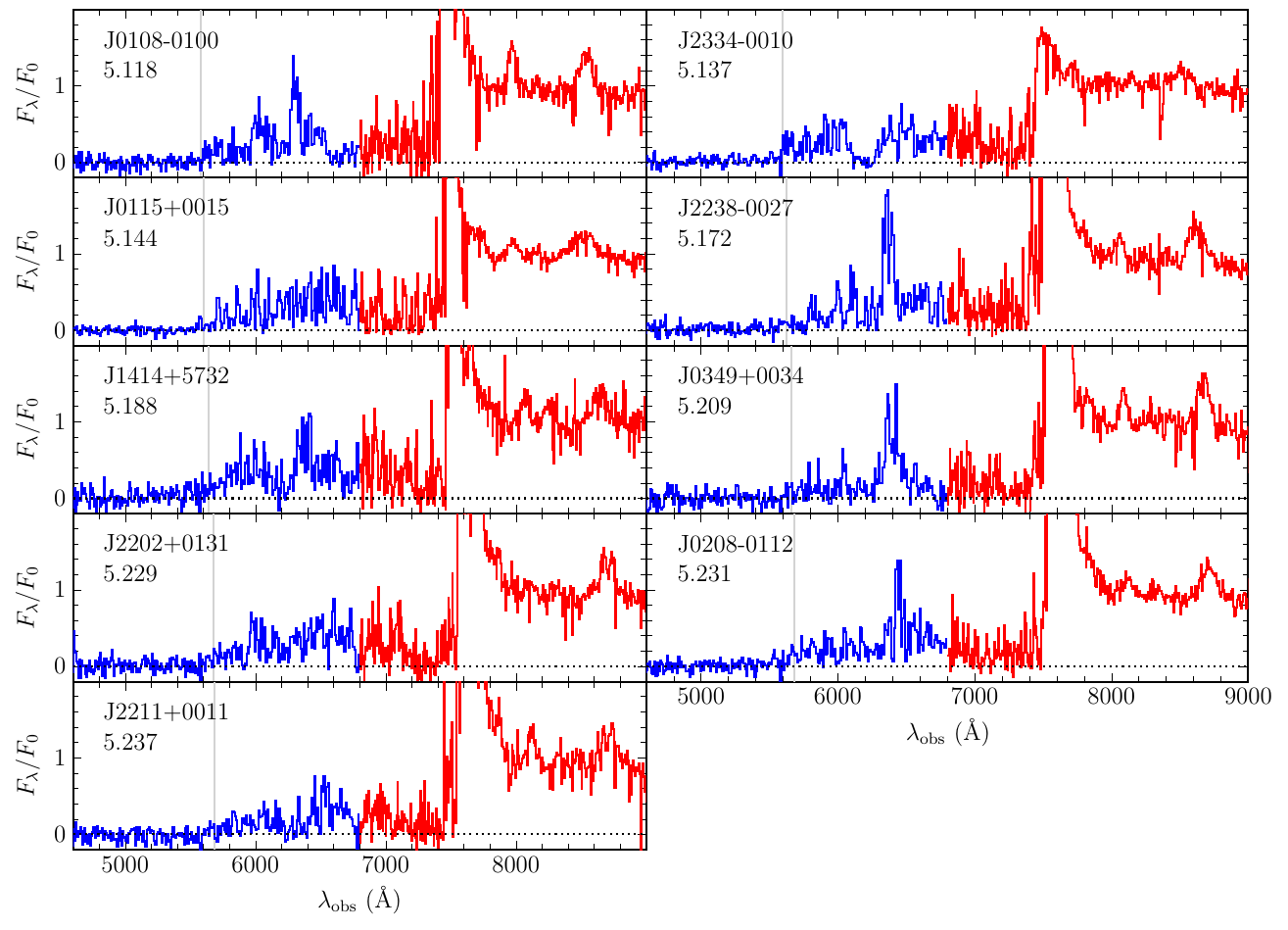}
   \vspace{-0.05in}
   \caption{LRIS spectra used in this work (continued from Fig.~\ref{fig:lris_spectra_1})}
   \label{fig:lris_spectra_2}
   \end{center}
   \end{minipage}
\end{figure*}

\begin{figure*}
   \centering
   \begin{minipage}{\textwidth}
   \begin{center}
   \includegraphics[width=1.0\textwidth]{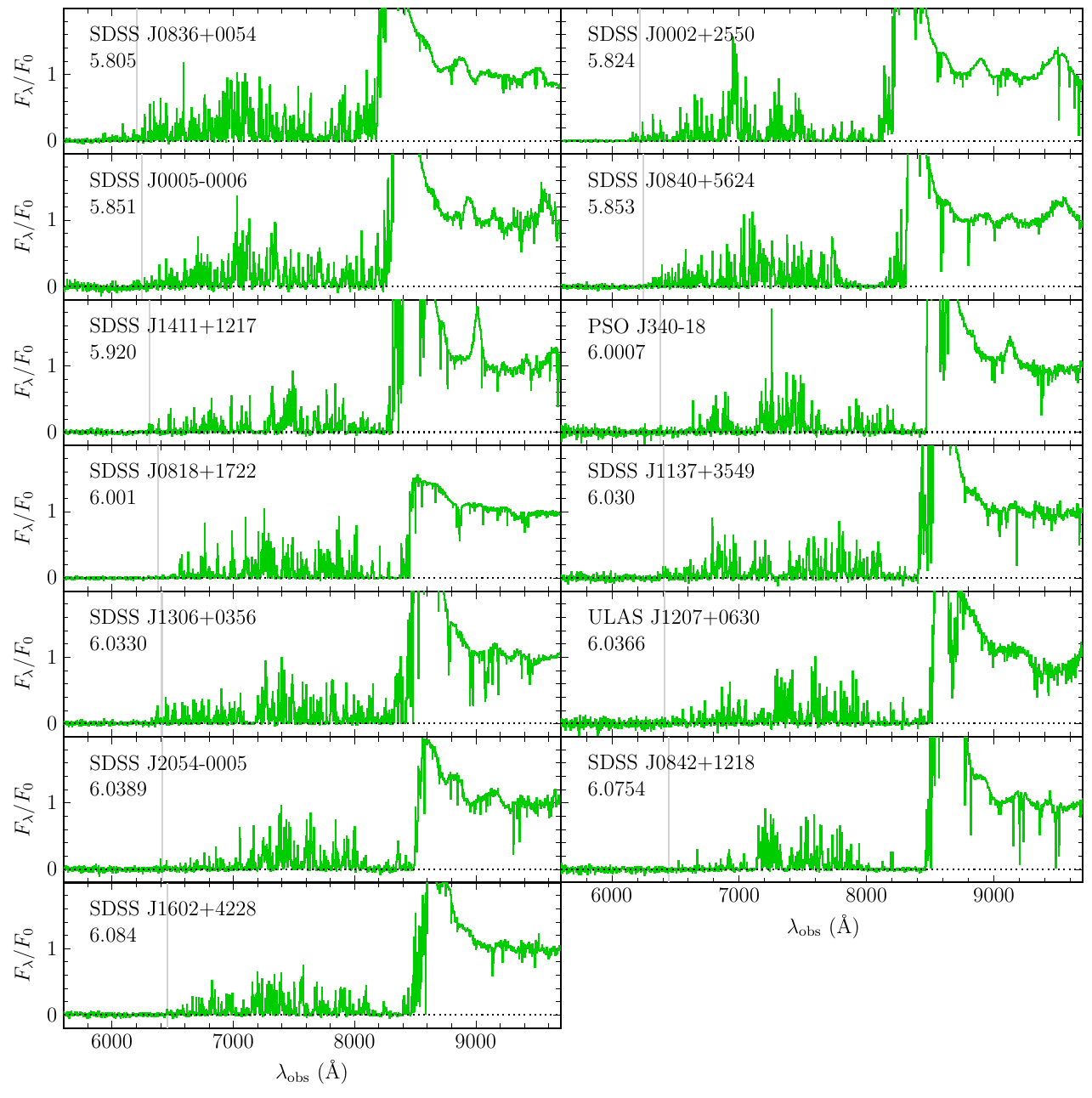}
   \vspace{-0.05in}
   \caption{ESI and X-Shooter spectra used in this work.  Panels are labeled with the QSO name and redshift.  For each QSO we plot flux per unit wavelength normalized by the continuum flux measured over rest-frame 1270--1380~\AA.  The spectra have been median filtered using a 3-pixel sliding window.  Vertical lines mark the Lyman limit wavelength in the rest frame of the QSO.}
   \label{fig:highz_spectra_1}
   \end{center}
   \end{minipage}
\end{figure*}

Here we present the individual spectra included in the LRIS and ESI + X-Shooter composites.  The LRIS spectra are plotted in Figs.~\ref{fig:lris_spectra_1} and \ref{fig:lris_spectra_2}.  The ESI and X-Shooter spectra are plotted in Fig.~\ref{fig:highz_spectra_1}.  In each case the spectra are normalized by the flux measured over rest-frame 1270--1380~\AA.  A vertical line marks the Lyman limit in the rest frame of the QSO.  For examples of the individual GGG spectra, see \citet{worseck2014}.

\bsp	
\label{lastpage}
\end{document}